\documentclass[lettersize,journal]{IEEEtran}

\usepackage{lipsum}
\usepackage{soul} 
\usepackage{nicefrac}

\usepackage{tipa}

\usepackage{enumitem}

\usepackage{multicol}

\usepackage[utf8]{inputenc}
\usepackage[T1]{fontenc}
\usepackage{textcomp}
\usepackage{microtype}
\usepackage{calc}
\usepackage{xcolor} 
\usepackage{footnote}
\usepackage{csquotes}

\usepackage{pgfplots}
\usepackage{graphicx}
\usepackage{tikz}
\usepackage{subcaption}

\usepackage{mathtools}
\usepackage{amsmath,amsfonts,amssymb}
\usepackage{algorithmic}

\usepackage[flushleft]{threeparttable}
\usepackage{multirow}
\usepackage{array}
\usepackage{longtable,supertabular}
\usepackage{tabularx}
\usepackage{booktabs}
\usepackage{rotating} 
\usepackage{multirow}

\usepackage[per-mode=symbol-or-fraction,range-units=single,list-units=single,detect-all]{siunitx}
\DeclareSIUnit\beat{beats}
\DeclareSIUnit\decibelm{dBm}
\DeclareSIUnit\decibeli{dBi}
\usepackage{csquotes}
\usepackage[backend=biber,style=ieee,doi=false,isbn=false,mincitenames=1,maxcitenames=2]{biblatex}
\DeclareFieldFormat{sentencecase}{#1} 
\DeclareFieldFormat{titlecase}{#1} 
\usepackage{xpatch}
\xpatchbibmacro{textcite}{\addspace}{\addnbspace}{}{}
\xpatchbibmacro{Textcite}{\addspace}{\addnbspace}{}{}

\DefineBibliographyStrings{english}{
	andothers = et~al\adddot\addspace
}

\usepackage[hidelinks]{hyperref}
\usepackage[capitalise,noabbrev]{cleveref}
\Crefformat{figure}{#2Fig.~#1#3}
\Crefformat{equation}{#2Eq.~(#1)#3}

\pgfplotsset{compat=1.18}

\input{acro}

\begin{document}

\title{Rejuvenating the IRS: Researching the Policy for the Low Communication Overhead}
\title{Rejuvenating IRS: Toward Low Communication Overhead Policies}
\title{Rejuvenating IRS: AoI-based Low Overhead Reconfiguration Design}


\author{\IEEEauthorblockN{%
    Jorge Torres G\'omez\IEEEauthorrefmark{1},
    Joana Angjo\IEEEauthorrefmark{1},
    Moritz Garkisch\IEEEauthorrefmark{2},
    Vahid Jamali\IEEEauthorrefmark{3}, 
    Robert Schober\IEEEauthorrefmark{2},
    and Falko Dressler\IEEEauthorrefmark{1}}%
    \\
    \IEEEauthorblockA{\IEEEauthorrefmark{1}%
        School of Electrical Engineering and Computer Science, TU Berlin, Berlin, Germany%
    }\\
    \IEEEauthorblockA{\IEEEauthorrefmark{2}%
        Institute for Digital Communications, University of Erlangen-Nuremberg, Erlangen, Germany%
    }\\
    \IEEEauthorblockA{\IEEEauthorrefmark{3}%
        Department of Electrical Engineering and Information Technology, Technical University of Darmstadt, Germany%
    }%

}

\markboth{IEEE Transaction on Wireless Communications,~Vol.~XX, No.~YY, October~2024}%
{J. Torres Gómez \MakeLowercase{\textit{et al.}}: Rejuvenating IRS}


\maketitle

\begin{abstract}
\Ac{IRS} technologies help mitigate undesirable effects in wireless links by steering the communication signal between transmitters and receivers.
\Ac{IRS} elements are configured to adjust the phase of the reflected signal for a user's location and enhance the perceived \ac{SNR}.
In this way, an \ac{IRS} improves the communication link but inevitably introduces more communication overhead.
This occurs especially in mobile scenarios, where the user's position must be frequently estimated to re-adjust the \acs{IRS} elements periodically.
Such an operation requires balancing the amount of training versus the data time slots to optimize the communication performance in the link.
Aiming to study this balance with the \ac{AoI} framework, we address the question of how often an \acs{IRS} needs to be updated with the lowest possible overhead and the maximum of freshness of information.
We derive the corresponding analytical solution for a mobile scenario, where the transmitter is static and the \ac{MU} follows a random waypoint mobility model.
We provide a closed-form expression for the average \ac{PAoI}, as a metric to evaluate the impact of the \ac{IRS} update frequency.
As for the performance evaluation, we consider a realistic scenario following the IEEE 802.11ad standard, targeting the mmWave band.
Our results reveal that the minimum achievable average \ac{PAoI} is in the microsecond range and the optimal \ac{IRS} update period is in the seconds range, causing $\SI{9}{\percent}$ overhead in the link when the \ac{MU} moves at a velocity of $\SI{1}{\meter\per\second}$.
\end{abstract}

\begin{IEEEkeywords}
Age of information, intelligent reflecting surfaces, mobility, mmWave, WiFi.
\end{IEEEkeywords}

\acresetall
\section{Introduction}
\IEEEPARstart{R}{ealizing} \ac{IRS}-assisted communication links are considered a powerful paradigm to enhance next-generation wireless communication.
\acp{IRS}s allow reshaping the communication channel for improved \ac{QoS} through low-cost reflective patches organized as planar arrays \cite{wu2020towards}.
This technology enables the steering of communication signals by adjusting the phase introduced by the \ac{IRS} elements, enhancing \ac{SNR} levels in targeted regions.
Such a solution finds application in mmWave band networks, at scalable cost and power consumption levels~\cite{chen2023reconfigurable-intelligent-surface}.

\begin{figure}
 \centering
 \includegraphics[width=0.8\linewidth]{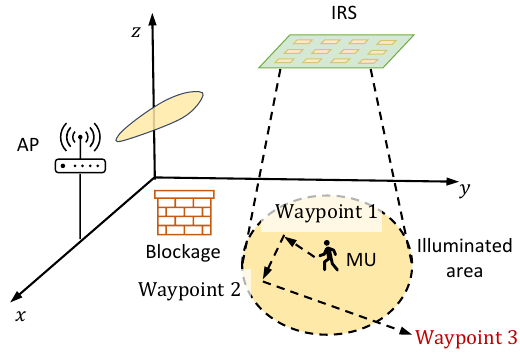}
 \caption{Illustration of the mobility of the \ac{MU} and \ac{IRS} coverage area.}
 \label{fig_location}
\vspace{-.8em}
\end{figure}

A key aspect of ensuring a given \ac{SNR} is the use of low-complexity methods to reconfigure the \ac{IRS} elements.
Examples include the partitioning of the \ac{IRS} array into multiple tiles \cite{najafi2021physicsbased}, the development of joint beam training and positioning for fast beam-tracking schemes \cite{wang2021joint}, and the use of variable-size codebooks \cite{jamali2021power}.
Besides, for the communication system to operate, the passive \ac{IRS} elements must be frequently updated using these methods while keeping track of the \acfp{MU} position in the network~\cite{hu2021twotimescale}.
Consequently, the \ac{IRS} should track the \ac{MU} position regularly and reconfigure the \ac{IRS} elements to illuminate the \ac{MU} with the new position.

We illustrate this requirement with the diagram depicted in \Cref{fig_location}, where an \ac{MU} is connected to an \ac{AP} whenever it is located within the illuminated area, as there is \ac{NLOS} link to the \ac{AP}.
As a result, the quality of the communication link in this mobile scenario is determined not only by the \ac{SNR} in the link but also by the overhead incurred each time the \ac{MU} is localized and the \ac{IRS} is updated, as during these time slots, data transmissions are interrupted.
The more frequently the \ac{IRS} is updated, the better the guarantees for the \ac{SNR}, but the fewer the opportunities for the users to transmit data, i.e., increased overhead.
On the contrary, the less frequently the \ac{IRS} is updated, the lower the communication link quality as the \ac{MU} might leave the illuminated area and experience a drop in \ac{SNR}.

In this scenario, a training protocol for the \ac{IRS} reconfiguration can be designed that accounts for the update period with the mobility of the \acp{MU} \cite{saggese2024impact}.
Reported examples in the literature evaluate the update period assuming the worst-case scenario, where a fixed channel coherence time is considered, as in \cite{cai2021downlink}, or by assuming the \ac{MU} deterministically moves to the edge of the coverage area and at a constant speed~\cite{laue2022performance}.
More straightforwardly, the authors in \cite{hashida2022adaptive} evaluate the \ac{IRS} reconfiguration period to maximize the link's sum rate.
%
The pilot interval to estimate the channel is increased (reducing overhead) until the impact of channel variability starts decrementing the \ac{SNR}.
While this study explicitly assesses the time configuration period, it examines the effect of \ac{MU} mobility solely through its velocity component, overlooking its specific mobility pattern.


In this paper, we adopt a more holistic approach utilizing the \ac{AoI} concept to evaluate the \ac{IRS} update period.
The \ac{AoI} concept encompasses all aspects of the communication link in a single formulation, including packet transmission periods, channel impairments, and delays within the communication pipeline~\cite{pappas2023age,lyu2022weighted}.
\Ac{AoI}-based solutions for \ac{IRS}-assisted links have been previously reported to maximize information freshness.
Existing research targets packet scheduling mechanisms in various network settings, e.g., in \ac{IRS}-empowered \ac{UAV} systems~\cite{samir2021optimizing,samir2021optimizing,fan2023risassisted}, \ac{SWIPT} \cite{lyu2023optimizing}, \ac{NOMA} networks \cite{feng2022optimizing}, and \ac{IRS}-assisted secure communications~\cite{wang2023intelligent}.
Nonetheless, these studies neglect the immediate effects of overhead from reconfiguration packets, particularly when balancing overhead and communication performance in mobile contexts.

We apply the \ac{AoI} in the mobile scenario depicted in \Cref{fig_AoI}, where a \ac{MU} does not have \ac{LOS} condition to an \ac{AP} and communication occurs through the \ac{IRS} only.
We research a reconfiguration strategy for a low-overhead transmission scheme focusing on minimizing the \ac{PAoI} metric.
We identify this approach as the missing component in today's literature for evaluating the optimal update period of the \ac{IRS} elements.
Pursuing this research direction, our main contributions can be summarized as follows:
\begin{itemize}
    \item We characterize the dynamic behavior of an \ac{IRS}-assisted link as a \ac{MRGP}, explicitly accounting for \ac{MU} mobility and the underlying transmission scheme.
    %
    \item We formulate the trade-off between performance and overhead in an \ac{AoI}-based framework.
    The proposed \ac{AoI} framework encompasses the dynamics of the \ac{IRS}-assisted link and can be used to determine the optimal periodicity for updating the \ac{IRS}.\footnote{This framework establishes the basis for the formulation and solution of optimization problems for the communication system design in future work.}
    %
    \item We propose a three-step methodology to exploit this theoretical model in a practical communication scenario.
    We examine the overhead for realistic WiFi communication links in the mmWave band, however, this methodology also applies to other communication scenarios.
\end{itemize}
%


The rest of this paper is organized as follows.
We discuss the reported research on the join topics \ac{IRS} and \ac{AoI} in \Cref{sec_related_work}.
We sketch the system model in \Cref{sec_SystemModel} detailing the transmission setup.
We model the link status model as a semi-Markov regeneration process in \cref{sec_semimarkov}.
The \ac{AoI}-based problem formulation is presented in \Cref{sec_AoI}, deriving analytical expressions to evaluate the minimum of the average \ac{PAoI} metric.
We assess the performance in \Cref{sec_results} using the derived analytical expressions.
In this section, we also illustrate the impact of overhead and mobility on the average \ac{PAoI} metric for a realistic WiFi link.\footnote{We provide open access to the code that evaluates the \ac{AoI} metrics in \ac{IRS}-assisted link in \url{https://github.com/tkn-tub/IRS_Rejuvenation}} 
Finally, we provide concluding remarks in \Cref{sec_Conclusion}.

%

\section{Related Work}
\label{sec_related_work}

\ac{AoI} has been used in \acp{IRS}-assisted networks to characterize the freshness of monitored sensor data.
Existing studies target various scenarios like \ac{IoT} networks in urban environments \cite{shi2023outage}, where power and latencies are predominant restrictions, and also \ac{UAV}-assisted scenarios \cite{fan2023risassisted,lyu2022weighted,jiang2023average}.
Furthermore, \ac{AoI} has also been used to maximize the freshness of data between two users in covert communication links as in \cite{wang2023intelligent,wang2022age}.

The considered systems comprise links between nodes and \acp{BS} or \acp{AP} in urban environments with one \cite{feng2022optimizing,wang2023intelligent,muhammad2023optimizing,wang2022age} or multiple \acp{IRS}s \cite{shi2023outage}.
More complex scenarios also include energy transfer from \acp{AP} to nodes using simultaneous transmitting and reflecting \acp{IRS}s \cite{xie2023performance}, as well as \ac{SWIPT} mechanisms \cite{zhang2023aoi,lyu2023optimizing}.
Other reports consider communication between \acp{UAV} and \ac{IoT} nodes through a fixed \ac{IRS} \cite{fan2023risassisted}.
Furthermore, towards enhancing flexibility during deployment, existing solutions have integrated \acp{IRS}s with \ac{UAV} to avoid the latency introduced by relay nodes at the \ac{UAV} \cite{lyu2022weighted,jiang2023average,sun2023aoi}.

The freshness of data is mostly optimized using the average \ac{AoI} metric (and less frequently using the average \ac{PAoI} as, e.g., in \cite{feng2022optimizing}).
The \ac{AoI} metric is used, e.g., in combination with communication schemes like \ac{HARQ} error-control protocols \cite{shi2023outage},
channel access mechanisms with \ac{NOMA} \cite{jiang2023average,xie2023performance}, and full-duplex communication links \cite{muhammad2023optimizing}.
Integrating \ac{UAV}-assisted links, the information freshness is evaluated readily with the average \ac{AoI} metric~ \cite{lyu2022weighted}, using the average of \ac{AoI} increase~\cite{fan2023risassisted}, and also as a multi-objective function using an \ac{URLLC}-based formulation, as in \cite{sun2023aoi}.


\Ac{AoI}-based solutions often come with certain restrictions.
The power level is typically part of the restriction due to the limited battery capabilities of \ac{IoT} nodes.
Reported research combines the communication reliability with the outage probability \cite{shi2023outage}, or the communication covertness limiting the detection probabilities of unauthorized users \cite{wang2023intelligent,wang2022age}. 
Further constraints include successful packet reception levels \cite{feng2022optimizing}, \ac{SNR} thresholds, the scheduling of the number of nodes per time slot, and the mobility range of \acp{UAV} \cite{lyu2022weighted,fan2023risassisted,samir2021optimizing,lyu2023optimizing,jiang2023average,sun2023aoi}.

Solutions using the \ac{AoI} formulation are developed both numerically and via \ac{ML}-based methods, aiming to control communication and mobility parameters jointly.
For instance, the alternating optimization algorithm in \cite{zhang2023aoi} is used to find the beamforming vectors, configure the \ac{IRS}, and the scheduling of nodes.
Genetic algorithms are also reported to optimize the power and time scheduling of nodes \cite{xie2023performance}.
\Ac{SCA} \cite{lyu2022weighted,wang2023intelligent} and the \ac{BCD} algorithms \cite{jiang2023average} are reported to jointly control \ac{UAV} mobility and the scheduling of \ac{IoT} nodes \cite{jiang2023average}, as well as to jointly configure the \ac{IRS} and the packet length \cite{wang2023intelligent}.
Using \ac{DRL}, off-on policies have been reported for various joint designs with reinforcement Q-learning, \ac{DQN}, and \ac{PPO} policies.
For instance, reinforcement Q-learning is reported to jointly find the location of the \ac{UAV}-\acp{IRS} \cite{sun2023aoi}.
\ac{IRS} and the packet service time are also jointly configured with \ac{DQN} \cite{fan2023risassisted}.
\acp{DQN} has also been proposed for the evaluation of the transmission scheduling only, as in \cite{sun2023aoi}.

Despite the rich literature and the optimal configuration of \acp{IRS}s, there is no solution for the optimal periodicity for reconfiguring the \ac{IRS} elements.
Analyzing the optimal periodicity allows determining the right balance between the overhead introduced for configuring the \ac{IRS} and the communication performance.
As we continue in the following sections, we will illustrate the relevancy of this concern in mobile scenarios, where the \ac{IRS} has to be frequently updated to guarantee a good communication performance, i.e., a particular \ac{QoS} level. 

%

\section{System Model}
\label{sec_SystemModel}
We study a communication downlink between an \ac{AP} and an \ac{MU} that moves at walking speed.
The communication link is facilitated via the reflection of an \ac{IRS}, where we assume \ac{NLOS} conditions between the \ac{AP} and the \ac{MU}, as depicted in \Cref{fig_location}.
The \ac{IRS} provides coverage for a circular area of radius $r_\mathrm{in}$, a shape that has also been considered in wireless scenarios, as in~\cite{segata2023enabling}.\footnote{We assume the circular geometry for the ease of evaluating the impact of \ac{MU} mobility, as introduced later in \Cref{eq_first_hitting}.
We remark that this assumption does not limit the general model introduced in \Cref{sec_semimarkov}, which is independent of the geometry of the illuminated area.}
The achieved \ac{SNR} is sufficiently high in the illuminated area to realize error-free transmissions with a certain data rate.
Under this general model, three important elements of the communication link include:  i) the \ac{IRS}-assisted link abstracted with the \ac{SNR} in the illuminated area, ii) the transmission frame in the downlink delimiting the time slots for communication, and iii) the communication-link dynamics given the \ac{MU} mobility model.
We elaborate on these elements in the following subsections.

\subsection{\ac{IRS}-assisted Link Model}
\label{sec_IRS}

We model the impact of the \ac{IRS}-assisted link using the \ac{SNR} in the \ac{MU} plane.
At a given observation point in the $xy$ plane, denoted as $\mathbf{p}_\mathrm{obs}$, the \ac{SNR} is given by the relation
\begin{equation}\label{eq_SNR}
    \mathrm{SNR}(\mathbf{p}_\mathrm{obs})=|g_\mathrm{IRS}(\mathbf{p}_\mathrm{obs})|^2\frac{P_\mathrm{Tx}\mathrm{PL}_\mathrm{AP,IRS}\mathrm{PL}_{\mathrm{IRS},\mathbf{p}_\mathrm{obs}}}{P_\mathrm{N}},
\end{equation}
which follows from \cite[Eq. (2)]{najafi2021physicsbased}, where $P_\mathrm{Tx}$ is the transmit power, $P_\mathrm{N}$ is the noise power, the variable $\mathrm{PL}_\mathrm{AP,IRS}$ is the free-space pathloss \ac{AP}-to-\ac{IRS} link, given by
\begin{equation}
    \mathrm{PL}_\mathrm{AP,IRS}=\left(\frac{\lambda}{2\pi||\mathbf{p}_\mathrm{IRS}-\mathbf{p}_\mathrm{AP}||}\right)^2,    
\end{equation}
where $\lambda=c/f$ is the wavelength of the transmitted signal, $c$ is the speed of light, and $f$ the center frequency for transmission.
Besides, $\mathbf{p}_\mathrm{IRS}$ denotes the position of the \ac{IRS}, $\mathbf{p}_\mathrm{AP}$ refers to the position of the \ac{AP}, and $||\cdot||$ is the norm of a vector.

In \cref{eq_SNR}, $\mathrm{PL}_{\mathrm{IRS},\mathbf{p}_\mathrm{obs}}$ is the free-space pathloss between the \ac{IRS} and the observation point $\mathbf{p}_\mathrm{obs}$ (within the \ac{MU} mobility plane), given by
\begin{equation}
    \mathrm{PL}_{\mathrm{IRS},\mathbf{p}_\mathrm{obs}}=\left(\frac{\lambda}{2\pi||\mathbf{\mathbf{p}_\mathrm{obs}-p_\mathrm{IRS}}||}\right)^2.  
\end{equation}
Furthermore, $g_\mathrm{IRS}(\mathbf{p}_\mathrm{obs})$ is the \ac{IRS} gain, evaluated as the sum of gains for each individual \ac{IRS}-element, given by
\begin{align}\label{eq_gain_IRS}
    g_\mathrm{IRS}(\mathbf{p}_\mathrm{obs})=\tilde{g}\,\sum_{n=1}^N\mathbb{I}&_{\left\{||\mathbf{p}_n-\mathbf{p}_\mathrm{IRS}||<\frac{L}{2}\right\}}\times\\
    &\times e^{j\frac{2\pi}{\lambda}(||\mathbf{p}_\mathrm{n}-\mathbf{p}_\mathrm{AP}||+||\mathbf{p}_\mathrm{obs}-\mathbf{p}_n||)}e^{j\omega_\mathrm{IRS}(n)}, \nonumber
\end{align}
which follows from \cite[Eq. (4)]{alexandropoulos2022nearfield}, where $N$ is the total number of \ac{IRS} elements, $L$ is the radius of the circumference inscribed in the \ac{IRS} plane (the purpose of this parameter is elaborated later in this section), $\omega_\mathrm{IRS}(n)$ is the \ac{IRS} phase shift corresponding to its $n$-th element, $\mathbf{p}_\mathrm{n}$ is the position of the $n$-th \ac{IRS} element,  and $\tilde{g}$ denotes the maximum gain of a given \ac{IRS} element, evaluated as
\begin{equation}
    \tilde{g}=\frac{\sqrt{4\pi}d_wd_h}{\lambda};
\end{equation}
see \cite[Eq. (12)]{najafi2021physicsbased}, where $d_w$ and $d_h$ refer to the width and the height of the unit cell, respectively.

In \Cref{eq_SNR}, the \ac{IRS} phase shift coefficients are the primary design variables, while the remaining parameters are predetermined by the scenario setup, as illustrated in \Cref{fig_location}.
The value of $\omega_\mathrm{IRS}(n)$ is obtained based on the design proposed in \cite[Sec. III. B 2)]{jamali2022lowto-zero-overhead} and \cite[Sec. III A]{alexandropoulos2022nearfield} to illuminate with a given \ac{SNR} the widest area possible. 
This coefficients design realizes a linear mapping function; see \cite[Eqs. (6) and (7)]{jamali2022lowto-zero-overhead}, which focuses each radiation point from the \ac{IRS} into an arbitrary rectangular area in the $xy$ plane.
The size of the projected rectangular area is adjusted with the beam-width parameters $\Delta x$ and $\Delta y$ in \cite[Eq. (6)]{alexandropoulos2022nearfield} and \cite[Eq. (7)]{jamali2022lowto-zero-overhead}.
However, to generate a circular illuminated area, we programmatically disconnect those \ac{IRS} elements that are not contributing to the circular shape; that is, the corresponding coefficients are set to zero in \eqref{eq_gain_IRS} with the indicator function $\mathbb{I}$.
The indicator function is equal to $1$, whenever the position of the \ac{IRS} element is within the inscribed circle of the \ac{IRS}, as follows from the condition $(||\mathbf{p}_n-\mathbf{p}_\mathrm{IRS}||<\frac{L}{2})$ \footnote{Hence, the activated \ac{IRS}-elements resemble an \ac{IRS} with circular shape.
The indicator function is equal to $1$, whenever the position of the \ac{IRS} element is within the inscribed circle of the \ac{IRS}, as follows from the condition $(||\mathbf{p}_n-\mathbf{p}_\mathrm{IRS}||<\frac{L}{2})$.
We note that we adopt this strategy to have a simple mapping function.
In principle, more sophisticated mapping functions or optimization-based designs can be developed to exploit the disconnected \ac{IRS} elements.}.
We provide an illustrative example for this calculation in \Cref{sec_scenario}, where a WiFi scenario setup is evaluated.\footnote{The Matlab code for the \ac{IRS} coefficient implementation is provided in \url{https://github.com/tkn-tub/IRS_Rejuvenation}}


\subsection{Transmission frame}

\begin{figure}
 \centering
 \includegraphics[width=0.8\linewidth]{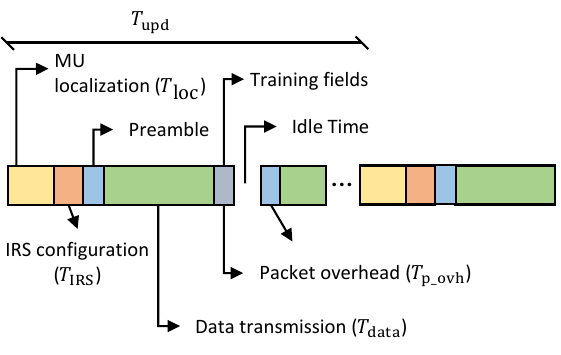}
 \caption{Frame structure for \ac{IRS}-assisted downlink transmission~\cite{jamali2022lowto-zero-overhead}.}
 \label{fig_transmission}
\vspace{-.8em}
\end{figure}

Transmissions between the \ac{AP} and the \ac{MU} take place according to the frame structure depicted in \Cref{fig_transmission}, which follows from \cite{jamali2022lowto-zero-overhead}.
Communication is performed in three phases.
Firstly, the \ac{MU} is localized by the \ac{AP} along the time interval~\mbox{$T_\mathrm{loc}$}, as depicted within the first slot in \Cref{fig_transmission}.
Using a given localization algorithm, the user is localized within the outer circle in \Cref{fig_location}.\footnote{Localization algorithms that do not rely on \ac{IRS} are surveyed in \cite{zafari2019survey}, whereas \ac{IRS}-assisted localization schemes are reviewed in \cite{zhang2023multiperson}.
The design and analysis of the methodology proposed in this paper are independent of the specific adopted localization scheme.}
Secondly, the \ac{IRS} is configured in the time interval $T_\mathrm{IRS}$ to illuminate the \ac{MU}.
Thirdly, data transmission occurs per packet, where each packet comprises a preamble and a training field, as additional overhead along the interval~$T_\mathrm{p\_ovh}$, and the data in time interval $T_\mathrm{data}$.
Following this transmission scheme, the \ac{MU} is localized, and the \ac{IRS} reconfigured with the period~$T_\mathrm{upd}$ when the \ac{MU}'s position is again updated, and the \ac{IRS} reconfigured. 
Implicitly, we also assume there is always data to transmit in the available slots, and queues are not included as we consider a point-to-point link only.

Using this transmission scheme, the data communication period increases with the overhead introduced in time slots $T_\mathrm{loc}$, $T_\mathrm{IRS}$, $T_\mathrm{p\_ovh}$ and the idle time $(T_\mathrm{idle})$.
The resulting overhead along the time interval $T_\mathrm{upd}$ can be evaluated as
\begin{equation}\label{eq_ovh}
T_\mathrm{ovh}=T_\mathrm{loc}+T_\mathrm{IRS}+c_h(T_\mathrm{p\_ovh}+T_\mathrm{Idle}),
\end{equation}
where
%
\begin{equation}\label{eq_ch}
c_h=\left\lceil\frac{T_\mathrm{upd}-(T_\mathrm{loc}+T_\mathrm{IRS})}{T_\mathrm{p\_ovh}+T_\mathrm{data}+T_\mathrm{Idle}}\right\rceil, 
\end{equation}
is the number of packets emitted during~$T_\mathrm{upd}$, and $\left\lceil\cdot \right\rceil$ denotes the ceiling operation.
Due to this overhead, the \ac{AP} in the link can perform transmissions with the equivalent time
\begin{equation}\label{eq_T_Tx}
    T_\mathrm{Tx}=T_\mathrm{data}+\frac{T_\mathrm{data}}{T_\mathrm{upd}-T_\mathrm{ovh}}T_\mathrm{ovh}.
\end{equation}
The calculation in \Cref{eq_T_Tx} accounts for the impact of overhead on the effective data transmission time.
The second term in the sum above proportionally distributes the total overhead (evaluated with $T_\mathrm{upd}$) among the total of data packets fitting within the time interval $T_\mathrm{upd}$.
This is evaluated as we divide $T_\mathrm{ovh}$ by the number of packets within $T_\mathrm{upd}$, yielding $\frac{T_\mathrm{upd}-T_\mathrm{ovh}}{T_\mathrm{data}}$.
In practice, the \ac{AP} performs a new packet transmission at periods larger than $T_\mathrm{data}$ and given by $T_\mathrm{Tx}$.

\subsection{Mobility model for the {MU}}

As for the user's mobility, we assume the \ac{MU} moves according to the \ac{RWP} mobility model~\cite{bettstetter2004stochastic}.
In this model, the \ac{MU} repeatedly selects a random destination point in a given area $A$ (in our case, a circle of radius $r_\mathrm{out}>r_\mathrm{in}$, as depicted in \Cref{fig_location}) and moves with constant speed on a straight line between the departure and destination points.
The random selection of the destination point follows a uniform probability distribution in the area.
We also consider the case where the \ac{MU} stops for a random time interval when reaching a destination and the case when the user moves at a random speed to the destination as well.
We provide further details about the mobility models in \Cref{eq_first_hitting,sec_validation}.

\section{Modeling the Link Status as a Semi-Markov Regenerative Process}
\label{sec_semimarkov}

Before formulating the average \ac{PAoI}, we provide insights into evaluating the fraction of time that the communication link is in any of the following three states:
\begin{enumerate}[label=\roman*)]
\item Error-free transmission: This occurs when the \ac{MU} is located in the illuminated area, i.e., inside the inner circle in \Cref{fig_location}.
\item Transmission with error: This happens when the \ac{MU} moves out of the illuminated region.
\item Interruption of transmissions: Occurring when the \ac{IRS} is being reconfigured.
\end{enumerate}
These three states define all possible link statuses between the \ac{AP} and the \ac{MU}.
We evaluate the fraction of time for each given state as the ratio between the duration of a given state to the update interval~$T_\mathrm{upd}$, yielding the portion of $T_\mathrm{upd}$ where the link is error-free, transmission with errors, or interrupted.
Evaluating this fraction of time lets us to determine the average amplitude of the peaks in the \ac{AoI} curve and, accordingly, the average \ac{PAoI} that we formulate later in \Cref{sec_AoI}.

The transition time instants between the above three states are generally a result of a random process.
The \ac{MU} stays in the illuminated area for a random amount of time as long as its corresponding waypoints are inside the inner circle in \Cref{fig_location}.
However, depending on the periodicity of the \ac{IRS} reconfiguration ($T_\mathrm{upd}$), the \ac{MU} will reach the outside area more or less quickly.
It can also perform transitions between both areas due to the random mobility.

These transition patterns between areas are repeatedly generated during the \ac{MU} localization phase, again starting the process regularly.
The communication link will remain in a given state for a random time, except during the localization and \ac{IRS} reconfiguration, where the link is interrupted (third state above) for a fixed amount of time $(T_\mathrm{loc}+T_\mathrm{IRS})$; see the transmission scheme in \Cref{fig_transmission}.

We model the dynamics of these three states by an \ac{MRGP} \cite[Sec. 10.6]{cinlar1975introduction}, allowing us to evaluate the time intervals of these states.
In contrast to a pure Markov process, the \ac{MRGP} model introduces random durations among transitions between states, allowing us to account for the random transition between error-free transmission and transmission with errors in the communication link.
Besides, the \ac{MRGP} model defines regeneration points, defined as those time instants where the random process replicates.
In our case, the regeneration points are determined by the \ac{IRS} reconfiguration, which will place the \ac{MU} again inside the illuminated area, and the process repeats.

In contrast to a continuous-time Markov process, a \acl{MRGP} relaxes the requirement of time independence.
This requirement states that the waiting time for the next transition is independent of the time elapsed in the current state, which does not apply to the three states of the considered communication link; see \cite[Eq. (12.1.13)]{howard1971dynamic}.
In our current setup, the waiting time during \ac{MU} localization and \ac{IRS} reconfiguration is deterministic, while for the other two states (error-free transmission and transmission with error), the waiting time is random.

With the \ac{MRGP} model, we follow a similar procedure as in~\cite{garg1995analysis}, where a software system is periodically reset as preventive maintenance to subside failures.
Analogously, in our case, the software components refer to the process running at the \ac{AP} to implement the transmission scheme with the \ac{MU}, i.e., localizing the \ac{MU}, configuring the \ac{IRS}, and performing transmissions.
Failures are identified whenever the \ac{MU} is not in the illuminated area, and maintenance corresponds with the \ac{MU} localization and the \ac{IRS} reconfiguration, after which the \ac{MU} is again illuminated.
This way, the communication process resembles a fault-tolerant system as in \cite{garg1995analysis}.


To derive the \ac{MRGP} model, we follow the procedure in~\cite{garg1995analysis}, where the evolution of the communication link is described first with a \ac{MRSPN} scheme.
Then, we derive the reachability graph yielding the \ac{MRGP} model representation, see also \cite{choi1994markov}.
Finally, we evaluate the corresponding transition probabilities using the formulation of local and global kernels.

%

\subsection{Stochastic Petri Net Representation}

\begin{figure}
 \centering
 \includegraphics[width=0.8\linewidth]{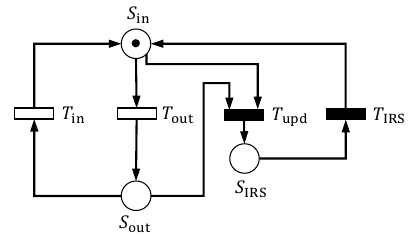}
 \caption{Pictorial representation of the \ac{MRSPN}.}
 \label{fig_petri_net}
\vspace{-0.8em}
\end{figure}

As depicted in \Cref{fig_petri_net}, the \ac{PN} representation allows for a graphical depiction of the flow of events~\cite{murata1989petri} and also identifies the transition probabilities and regeneration points of the \ac{MRGP} model \cite{choi1994markov}.
As shown in \Cref{fig_petri_net}, this representation ecompases the following features:
\begin{itemize}
    \item The link status: Denoted with the three circles, also refered to as places, where $S_\mathrm{in}$ stands for the event when the node is in the illuminated area (error-free transmissions), $S_\mathrm{out}$ indicates that the node is outside the illuminated area (transmissions with errors), and $S_\mathrm{IRS}$ denotes the case when the node is localized and the \ac{IRS} is reconfigured (no transmission happens).
    \item Transition between states: This is denoted with boxes, where the empty ones represent the random transitions between the states, while the filled boxes indicate the deterministic transitions. 
    \item Connections between states and transitions: These connections are represented with directed arrows, providing direction among states.
    \item The current state of the link: This is denoted with a token, depicted with a dot.
    \Cref{fig_petri_net} depicts the initial condition in the link with the dot in the circle $S_\mathrm{in}$, i.e., the node is in the illuminated area, and transmissions are error-free.
\end{itemize}
The boxes govern transitions between the circles (places) through the \textit{firing} of events.\footnote{Within the Petri Net representation places refers to states, i.e., $S_\mathrm{in}$, $S_\mathrm{out}$, and $S_\mathrm{IRS}$ in \Cref{fig_petri_net}.}
This firing refers to the boxes’ activation; upon activation, the token at the input is moved to the box’s output, representing the dynamic transition from the input to the output place.
Furthermore, a box only enables the token’s transition to the next state if a token is present at its input; otherwise, the firing does not occur. 
This representation also assumes that deterministic firing has priority over random firing. Whenever \( T_\mathrm{upd} \) occurs, the token transitions from place \( S_\mathrm{in} \) or \( S_\mathrm{out} \) to \( S_\mathrm{IRS} \).

As the transition is governed by the \ac{MU} location, the firing is accordingly defined with the following variables:
\begin{itemize}
    \item $T_\mathrm{out}$, a continuous stochastic variable indicating the \ac{MU} transition from the illuminated area to the outer area in \Cref{fig_location};
    \item $T_\mathrm{in}$, a continuous stochastic variable indicating the \ac{MU} transition from the outer area to the illuminated area;
    \item $T_\mathrm{upd}$, a deterministic variable indicating the \ac{MU}'s position is updated while using a given localization algorithm (running at the \ac{AP});
    this firing occurs at regular time intervals;
    \item $T_\mathrm{IRS}$, a deterministic variable indicating the \ac{IRS} has been reconfigured and transmissions start again.
\end{itemize}


Given the firing rules, the evolution of the \ac{PN} is described by the distribution of tokens at places $S_\mathrm{in}$,  $S_\mathrm{out}$, and $S_\mathrm{IRS}$.
This distribution of tokens is defined by a vector, referred to as a mark~\cite{peterson1977petri}, where each vector component is an integer number denoting the number of tokens at places $S_\mathrm{in}$, $S_\mathrm{out}$, and $S_\mathrm{IRS}$. 
In the network in \Cref{fig_petri_net}, there will be only three different marks as there are only a single token and three places.
These marks are evaluated as follows: Departing from the initial mark, we have the vector $M_\mathrm{IRS}=[1,0,0]$, denoting a single token is located at $S_\mathrm{IRS}$.
This mark denotes the case when the node is localized, and the \ac{IRS} reconfigured.
We evaluate the other two marks as $M_\mathrm{in}=[0,1,0]$, indicating a token is located at $S_\mathrm{in}$, and $M_\mathrm{out}=[0,0,1]$ indicating the token is located at $S_\mathrm{out}$.
These two cases imply that the \ac{MU} is within and outside the illuminated area, respectively.

\begin{figure}
 \centering
 \includegraphics[width=0.5\linewidth]{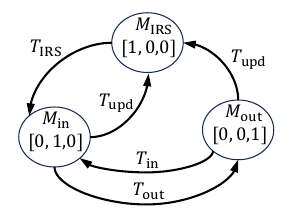}
 \caption{Reachability graph of the \ac{MRSPN}.}
 \label{fig_markov_chain}
\vspace{-.8em}
\end{figure}

Furthermore, the transitions of the token from one place to another can be represented by transitions between markers, which define the reachability graph as depicted in \Cref{fig_markov_chain}.
This graph indicates which mark is reachable from the others according to the random ($T_\mathrm{in}$, $T_\mathrm{out}$) and deterministic ($T_\mathrm{upd}$, $T_\mathrm{IRS}$) firing of events.
As a result of the process dynamics, all the marks are reachable from each other with a direct transition, except for $M_\mathrm{IRS}$, which always transits to $M_\mathrm{in}$.

%

\subsection{Modeling the Link Dynamics using MRGP}

The marks of the \ac{MRSPN} readily define an \ac{MRGP}; see \cite[Theorem 2]{choi1994markov}. 
The state space of the underlying \ac{MRGP} is directly given by the marks set as $\Omega=\{M_\mathrm{IRS},M_\mathrm{in},M_\mathrm{out}\}$.
The regenerative points are defined by the transitions to the state $M_\mathrm{IRS}$, where the stochastic process restarts again with the \ac{MU} mobility pattern.
Consequently, the regenerative set comprises only a single element namely $\Omega'=\{M_\mathrm{IRS}\}$.

The \ac{MRGP} is completely described by the local and global kernels probabilities (see \cite[Theorem 3]{choi1994markov}), which are defined as follows \cite[Definition 2]{garg1995analysis}.

\textbf{Global kernel} $\textbf{K}(t)$: This is an $m\times m$ matrix, where $m$ is the cardinality of the regenerative set $\Omega'$ (in our case $m=1$).
    The global kernel evaluates the process at the regeneration points (given by $T_\mathrm{upd}$) with the conditional probability
    \begin{align}\label{eq_global_kernel}
        \textbf{K}(t)=&[K_{11}(t)]=\,[\mathrm{Pr}\{Z(T_\mathrm{upd})=M_\mathrm{IRS},T_\mathrm{upd}\leq t\,|\,\nonumber\\ 
        &\hspace{5.0em} |Z(T_0)=M_\mathrm{IRS}\}],\\
        =\,&[u(t-T_\mathrm{upd})],\nonumber
    \end{align}
    where the Kernel's index ``\num{11}'' refers to state $M_\mathrm{IRS}$, $Z$ indicates the \ac{MU}'s state, $T_0$ denotes the initial time instant when the stochastic process starts, and $u(t)$ is the unit step function.
    We evaluate this conditional probability with the step function in \eqref{eq_global_kernel}, as the token is forced to return to place $S_\mathrm{IRS}$ at $T_\mathrm{upd}$, and the condition $Z(T_\mathrm{upd})=M_\mathrm{IRS}$ will occur with probability one whenever $t\geq T_\mathrm{upd}$, see \Cref{fig_petri_net}.
    In the meantime, the link is transitioning between $S_\mathrm{in}$ and $S_\mathrm{out}$.
    
\textbf{Local kernel} $\textbf{E}(t)$: This is an $m\times n$ matrix, where~$n$ is the cardinality of the state space $\Omega$ (in our case~\mbox{$n=3$}), that evaluates the behavior of the marks between two consecutive regenerative points.
    $\textbf{E}(t)$ results in the $1\times 3$ vector
    \begin{align}\label{eq_local_kernel}
        \textbf{E}(t)&=[E_{iq}(t)]=[\mathrm{Pr}\{Z(t)=Z_q,T_\mathrm{upd}> t\,|\, Z(T_0)=Z_i\}],\nonumber\\
        &=\left[
        \begin{matrix}
            E_{11}(t) & E_{12}(t) & E_{13}(t)     
        \end{matrix}
        \right].
    \end{align}
    where the lettered subscripts $i$ and $q$ denote the departing and arriving states, respectively, $i,\,q\,\in\{1,\,2,\,3\}$, and the corresponding $Z$'s are $Z_1=M_\mathrm{IRS}$, $Z_2=M_\mathrm{in}$, and \mbox{$Z_3=M_\mathrm{out}$}.
    In \Cref{eq_local_kernel}, the entry $E_{11}(t)$ refers to the probability that the process running at the \ac{AP} is in the localization and \ac{IRS} configuration phases (denoted as $M_\mathrm{IRS}$), given that the process always initiates at the same state $M_\mathrm{IRS}$.
    This probability is readily obtained as
    \begin{equation}\label{eq_E_11}
        E_{11}(t)=1-u(t-T_\mathrm{conf}),
    \end{equation}
    with
    \begin{equation}\label{eq_T_setting}
        T_\mathrm{conf}=T_\mathrm{loc}+T_\mathrm{IRS}.
    \end{equation}
    This formulation states that the \ac{MU} will remain with probability one in the $M_\mathrm{IRS}$ state along time interval $[0,\,T_\mathrm{conf}]$, whenever $T_\mathrm{upd}$ triggers.
    
    The term $E_{12}$ evaluates the probability that the process continues to be in $M_\mathrm{in}$ before the first regeneration point~$T_\mathrm{upd}$.
    We recall that $M_\mathrm{IRS}$ denotes the localization and \ac{IRS} configuration processes, after which the \ac{MU} is again illuminated. 
    Thereby, the token will be at place $S_\mathrm{in}$ immediately after $T_\mathrm{IRS}$ and will stay within the illuminated area with probability $\mathrm{P}_\mathrm{in}$ till the next regeneration point.
    Evaluating these two conditions together yields
    \begin{equation}\label{eq_E_12}
        E_{12}=\left\{\begin{array}{lr}
             0 & \text{when } t\leq T_\mathrm{conf}\\
             1 & \text{when } T_\mathrm{conf}<t\leq \frac{r_\mathrm{in}}{v}\\
             \mathrm{P}_\mathrm{in}(t-t_0<T_\mathrm{upd}) & \text{when } t> T_\mathrm{conf}
             \end{array}\right.
    \end{equation}
    where $\mathrm{P}_\mathrm{in}(t<T_\mathrm{upd})$ denotes the probability that the \ac{MU} remains inside the inner circle till the next regeneration point~$T_\mathrm{upd}$.
    In \Cref{sec_Pr_in} we evaluate the probability that the \ac{MU} leaves the illuminated circular area departing from an arbitrary location within the circle, here denoted with $\mathrm{P}_\mathrm{in}(t<T_\mathrm{upd})$. 
    As we assume that the localization phase centers the circle of the illuminated area in the \ac{MU} position, see \Cref{fig_location}, we introduce the time shift $t_0=T_\mathrm{conf}-\frac{r_\mathrm{in}}{v}$ accounting for the initial \ac{MU}'s location in the circle center instead, where $\frac{r_\mathrm{in}}{v}$ accounts for the minimum time the \ac{MU} takes to travel from the the center of the illuminated area to the outside area.
    This probability ultimately depends on the \ac{MU}'s mobility pattern.
    We illustrate the calculation and the validation of this probability in \Cref{sec_Pr_in} for various \ac{RWP} mobility models.
    
    Similarly, $E_{13}$ is evaluated as follows
    \begin{align}\label{eq_E_13}
        E_{13}(t)=&\mathrm{P}_\mathrm{out}(t-T_\mathrm{conf}<T_\mathrm{upd})u(t-T_\mathrm{conf}),
    \end{align}
    %
    where $\mathrm{P}_\mathrm{out}(t<T_\mathrm{upd})$ denotes the probability that the \ac{MU} stays in the outer area and assuming the \ac{MU}'s initial position is in the same outer area.
    In this equation, we introduce the time shift $T_\mathrm{conf}$ as this event always occurs after the \ac{IRS} is reconfigured.
    We evaluate this probability in \Cref{sec_Pr_out}.

The two kernels in \eqref{eq_global_kernel} and \eqref{eq_local_kernel} allow us to evaluate the steady-state transition probability, as \cite[Eq. (5)]{garg1995analysis}
\begin{equation}\label{eq_trans_prob}
    \pi_q=\frac{\sum_{k\in \Omega} p_k\alpha_{kq}}{\sum_{k\in \Omega} p_k\sum_{l\in \Omega} \alpha_{kl}},
\end{equation}
where $q\in\,\{1,2,3\}$ denotes three link states, namely \ac{IRS} reconfiguration, error-free transmission, and transmision with errors, respectively.
Moreover, $\alpha_{k,q}$ is given by
\begin{equation}
    \alpha_{kq}=\int_0^\infty{E_{kq}(t)\mathrm{d}t},
\end{equation}
and $p_k$ in \Cref{eq_trans_prob} are the components of  $\textbf{p}$ as a $1\times n$ vector, which is the solution to the equation
\begin{equation}\label{eq_p}
    \textbf{p}=\textbf{p}\textbf{K}(\infty),
\end{equation}
under the condition $\sum_{i\in \Omega'}p_i=1$.
As this condition states that $p_1=1$ (there is only a single regeneration state as $M_\mathrm{IRS}$), and $\textbf{K}(\infty)=1$ (see \Cref{eq_global_kernel}), the evaluation of \eqref{eq_p} yields the trivial solution~\mbox{$\textbf{p}=[1\, 0\, 0]$}.
Replacing this solution into \eqref{eq_trans_prob} finally evaluates $\pi_j$ as
\begin{equation}\label{eq_trans_prob2}
    \pi_q=\frac{\alpha_{1,q}}{\sum_{l\in \Omega} \alpha_{1,l}},
\end{equation}

By definition, $\pi_q$ in \eqref{eq_trans_prob2} provides the proportion of time the token is on each place with respect to the time interval between regeneration points (see \cite[Theorem 6]{choi1994markov}).
In our case, it refers to the proportion of time the link is in either of the above-mentioned states: transmitting error-free, transmitting with errors, or non-transmitting.
We will use this equation to evaluate the average \ac{PAoI} in \Cref{sec_AoI}. 

\begin{figure}
\centering
{\includegraphics[width=0.8\linewidth]{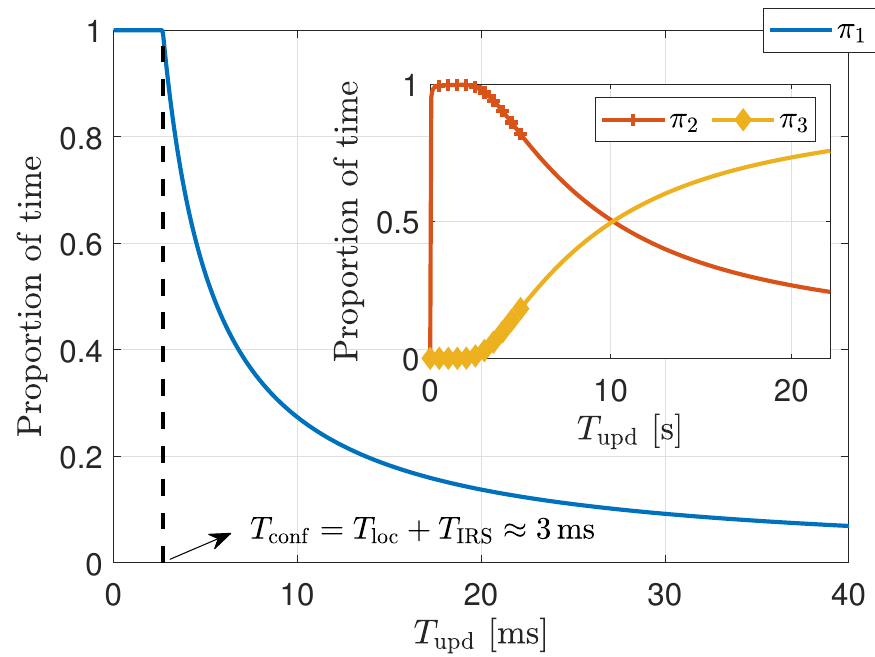}\label{fig_pi_2_3}}
\caption{Fraction of time (see \eqref{eq_trans_prob2}) for the three states in the link considering $r_\mathrm{in}=\SI{1.7}{\meter}$, $r_\mathrm{out}=\SI{3}{\meter}$, and mobility of constant speed $v=\SI{1}{\meter\per\second}$ without stop time.}
\label{fig_pi_vector}
\vspace{-.8em}
\end{figure}

The proportion of time vector $\pi_q$ will ultimately depend on the probability that the \ac{MU} is inside the illuminated area ($\mathrm{P}_\mathrm{in}$), outside ($\mathrm{P}_\mathrm{out}$), and the amount of time needed to localize and reconfigure the \ac{IRS} ($T_\mathrm{conf}$).
To illustrate, \Cref{fig_pi_vector} depicts the fraction of time for the three communication link states after numerically evaluating \eqref{eq_trans_prob2}, and using the calculations made in the next sections for the probabilities $P_\mathrm{in}$ and $P_\mathrm{out}$.
We choose to illustrate assuming the \ac{RWP} mobility model with speed $\SI{1}{\meter\per\second}$ for the \ac{MU} and with the inner radius of the illuminated area as $\SI{1.7}{\meter}$; this is the illuminated area where the \ac{IRS}-assisted link evaluates no packet errors (evaluated later in \Cref{sec_results}). 
The link is with probability one ($\pi_1=1$) in the localization and reconfiguration phases whenever $T_\mathrm{upd}\leq(T_\mathrm{loc}+T_\mathrm{IRS})$ and starts to decrease afterward.
As $T_\mathrm{upd}$ increases, the \ac{MU} reduces its probability to stay inside the illuminated area ($\pi_2$).
Eventually, the \ac{MU} starts to spend time in the outer area with increased probability, which follows the curve $\pi_3$.
In the following, we provide the calculations for $\mathrm{P}_\mathrm{in}$ with the \ac{RWP} mobility model.
A similar procedure will follow to evaluate~$\mathrm{P}_\mathrm{out}$.

\subsection{Evaluating the Probability for {MU} to Stay Inside the Illuminated Area}
\label{sec_Pr_in}

We evaluate the probability for error-free transmissions as the probability that the \ac{MU} stays inside the illuminated area (denoted as $\mathrm{P}_\mathrm{in}$ in \eqref{eq_E_12}).
To evaluate this probability, we first compute the first-hitting-time probability~\cite{redner2001guide} that the \ac{MU} reaches the outer area in \Cref{fig_location}. 
Denoting this probability as $P_\mathrm{fh\_out}(t)$, we readily
compute $\mathrm{P}_\mathrm{in}(t)$ as
\begin{equation}\label{eq_Pr_in}
    \mathrm{P}_\mathrm{in}(t)=1-P_\mathrm{fh\_out}(t).
\end{equation}
Following the \ac{RWP} mobility model, we decompose the calculation of $P_\mathrm{fh\_out}$ into two steps.
In the first step, we compute the probability for the \ac{MU} to leave the inner circle, see \Cref{fig_location}, with the first-hitting time probability.
We evaluate the probability for the number of consecutive waypoints inside the inner circle before the \ac{MU} transitions to the outer area.
In the second step, we evaluate the probability of the elapsed time between waypoints in the inner circle and add those random variables.
The sum provides a close estimate of the total time the \ac{MU} spends within the illuminated area, resulting in~$\mathrm{P}_\mathrm{in}(t)$.
We illustrate this approximation later with \Cref{sec_validation} and graphically in \Cref{fig_Pr_in_sim}.

\subsubsection{First-hitting time probability}
\label{eq_first_hitting}


To evaluate this probability, we need to compute the survival probability first; which evaluates the probability for the \ac{MU} to stay in the inner circle for $j$-consecutive waypoints.
Here denoted as $S_j$, this evaluation introduces the impact of the particular geometry of the illuminated area, in our case a circle of arbitrary radius~$r_\mathrm{in}$.
The probability of selecting a waypoint in the inner circle of area $A_\mathrm{in}$ is directly given as $p=\frac{A_\mathrm{in}}{A}$, where $A$ denotes the area of the outer circle in \Cref{fig_location}.
In this way, the value for $p$ inherently evaluates equal probability of displacing to any point in $A$, following the \ac{RWP} model; see \cite{bettstetter2004stochastic}.
Consequently, the probability that the \ac{MU} selects $j$-consecutive waypoints within the inner circle yields the survival probability \mbox{$S_j=\left(\frac{A_\mathrm{in}}{A}\right)^j,\, j\geqslant 0$}, as each waypoint selection is independently performed.
Given~$S_j$, the first-hitting probability is readily evaluated as \cite{redner2001guide}
%
\begin{align}\label{eq_first_time_discrete}
    P_{J\mathrm{out}}(j)&=S_J(j)-S_J(j+1)=\left(\frac{A_\mathrm{in}}{A}\right)^j\left(1-\frac{A_\mathrm{in}}{A}\right),\\
    &\frac{A_\mathrm{in}}{A}\!\in\!(0,\!1), \, j\geqslant 0, \nonumber
\end{align}
which represents the first-hitting time probability in the discrete space of jumps, and $J_\mathrm{out}$ is a random variable denoting the waypoint where the \ac{MU} is for the first time outside the illuminated area.




\subsubsection{First-hitting time probability in the time-space}
For each performed jump, the \ac{MU} moves a random distance $l_j$ with a given speed $v$, where the \ac{PDF} for $l_j$ is given as \cite[Eq. (23)]{bettstetter2004stochastic}
\begin{align}\label{eq_f_L}
    f_{L}(l_j)=\frac{8}{\pi r_\mathrm{in}}\frac{l_j}{2r_\mathrm{in}}\bigg(&\arccos\left(\frac{l_j}{2r_\mathrm{in}}\right)\\
    &-\frac{l_j}{2r_\mathrm{in}}\sqrt{1-\left(\frac{l_j}{2r_\mathrm{in}}\right)^2}\bigg). \nonumber
\end{align}
We evaluate \eqref{eq_f_L} considering the \ac{MU} displaces in the inner circle only, as we aim to evaluate the elapsed time interval before the \ac{MU} leaves it.
For this case, we assume $l_j\!\in[0,\,2r_\mathrm{in}]$, where  $r_\mathrm{in}$ denotes the radius of the inner circle in \Cref{fig_location}.

\begin{figure*}
\centering
\subfloat[\ac{RWP} mobility model when the \ac{MU} displaces with constant speed $\SI{1}{\meter\per\second}$, without stop time.]{\includegraphics[width=.32\textwidth]{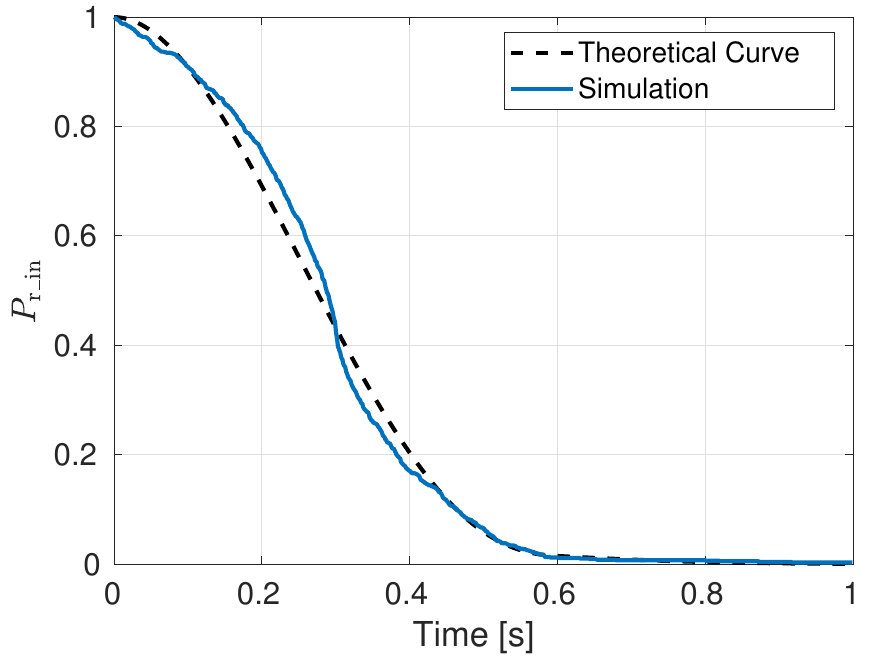}\label{fig_Pr_in_sim_RWP}}
\hfill
\subfloat[\ac{RWP} mobility model with random stop time following an exponential distribution $f_{T_p}(\tau)=\mu e^{-\mu \tau}$ and mean waiting time~\mbox{$\mu=\SI{2}{\second}$}.]{\includegraphics[width=.32\textwidth]{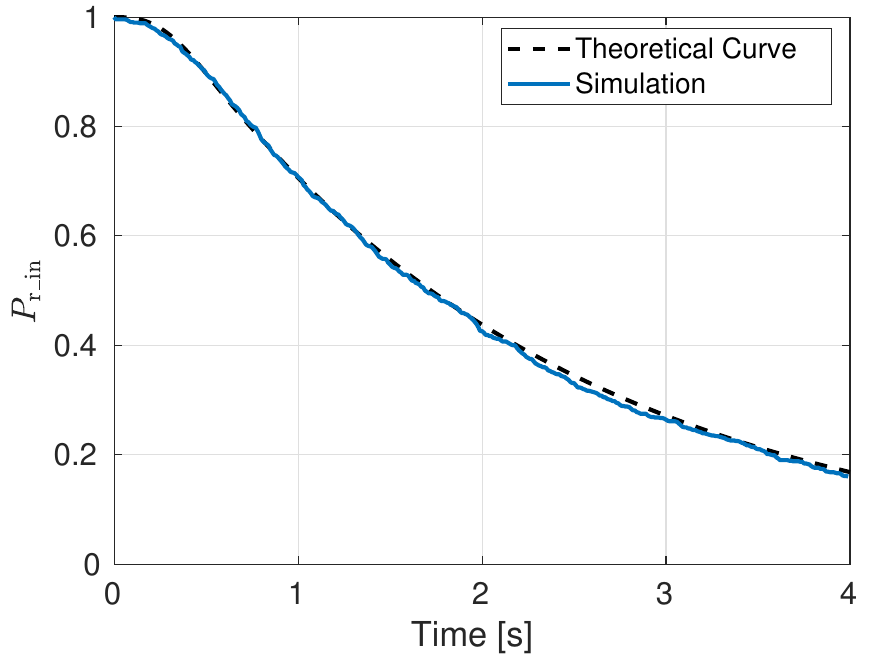}\label{fig_Pr_in_sim_RWP_stopTime}}
\hfill
\subfloat[\ac{RWP} mobility model when the \ac{MU} displaces with random speed based on the uniform distribution between~$v_\mathrm{min}$ and~$v_\mathrm{max}$, where~$v_\mathrm{min}=\SI{0.5}{\meter\per\second}$ and~$v_\mathrm{max}=\SI{1.5}{\meter\per\second}$.]{\includegraphics[width=.32\textwidth]{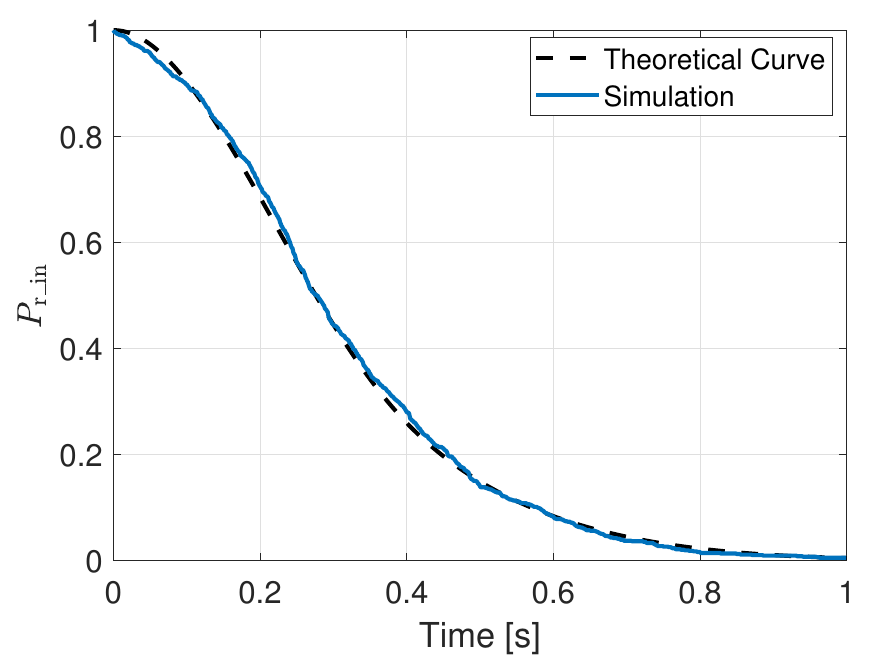}\label{fig_Pr_in_sim_RWP_random_time}}
\caption{Validation for the $P_\mathrm{in}$ in \eqref{eq_Pr_in} with simulation of the \ac{RWP} mobility for inner circle radius $\SI{1}{\meter}$. The exponential distribution in b) follows \cite[Example in page 561]{bettstetter2004stochastic}, the uniform distribution in c) follows \cite[Sec. 3.5.1]{bettstetter2004stochastic}.}
\label{fig_Pr_in_sim}
\vspace{-.8em}
\end{figure*}

Using the formulation in \Cref{eq_f_L}, we can consider three different cases for the \ac{MU} mobility pattern.
In the first case, the \ac{MU} moves with constant speed $v$ between waypoints along the time interval $\tau_j=\frac{1}{v}l_j$, where (see \cite[Eq. (28)]{bettstetter2004stochastic})
%
%
\begin{equation}\label{eq_f_tau_j_2}
    f_{\tau_j;r_\mathrm{in}}=vf_{L}(v\!\tau_j),\, \tau_j\!\in\!\left(0,\, \frac{2r_\mathrm{in}}{v}\right)
\end{equation}
%
In the second case, the \ac{MU} performs a random pause time $T_p$ with \ac{PDF} $f_{T_p}(t_p)$, yielding \cite[Eq. (39)]{bettstetter2004stochastic}
\begin{equation}\label{eq_f_tau_j_3}
f_{\tau'_j;r_\mathrm{in}}=\int_0^{\tau'}{f_{\tau_j;r_\mathrm{in}}(\tau)f_{T_p}(\tau'-\tau)\mathrm{d}\tau}.
\end{equation}
In the last case, the \ac{MU} displaces at a random speed $v$ following $f_V(v)$ as the \ac{PDF}, yielding \cite[Eq. (39)]{bettstetter2004stochastic}
\begin{equation}\label{eq_f_tau_j_4}
    f_{\tau'_j;r_\mathrm{in}}=\int_{v_\mathrm{min}}^{v_\mathrm{max}}{v}{f_{\tau_j;r_\mathrm{in}}(\tau)f_V(v)\mathrm{d}v}.
\end{equation}



We approximate the total time elapsed till the \ac{MU} displaces out of the inner circle as the sum of the time intervals along $J$-consecutive waypoints except for the last one as\footnote{This approximation is a pessimistic calculation for the elapsing time inside the illuminated area. 
The exact time is larger, and it evaluates the time interval just when the \ac{MU} intercepts the perimeter of the illuminated area.}
\begin{equation}\label{eq_tau_c}
    T_\mathrm{out}\approx\sum_{j=1}^{J_\mathrm{out}-1}\tau_j,
\end{equation}
where $J_\mathrm{out}$ is a random variable denoting the waypoint when the \ac{MU} leaves the inner circle for the first time, see for instance the case depicted in \Cref{fig_location} where $J_\mathrm{out}=3$.

To provide a formula for the probability that the \ac{MU} leaves the inner circle for the first time before $T_\mathrm{upd}$, we use the formula for the total probability, yielding \cite{papoulis2002probability}
\begin{equation}\label{eq_P_tau_j}
    P_\mathrm{fh\_out}(t)=\sum_{j=1}^\infty {P(T_\mathrm{out}|j)P_{J\mathrm{out}}(j)},
\end{equation}
as \eqref{eq_tau_c} is the sum of an arbitrary amount of random variables $\tau_j$, where $P(T_\mathrm{out}|j)$ denotes the probability for the total elapsed time along $j$-jumps, evaluated through the corresponding \ac{PDF}\footnote{As the elapsing time is the sum of independent time intervals between jumps (see \eqref{eq_tau_c}), its corresponding \ac{PDF} can be evaluated using the $N$-fold convolution of $f_{\tau_j}$; see \cite{papoulis2002probability}.}
\begin{equation}\label{eq_f_tau_c}
f_{T_\mathrm{out}}(t_\mathrm{out}|j)=
    \underbrace{f_{\tau_j;r_\mathrm{in}}\ast \cdots \ast f_{\tau_j;r_\mathrm{in}}}_{j}\big|_{\tau_j=t_\mathrm{out}},
\end{equation}
where $f_{\tau_j;r_\mathrm{in}}$ is obtained using \Cref{eq_f_tau_j_2,eq_f_tau_j_3,eq_f_tau_j_4} depending on the mobility model.\footnote{Evaluating \eqref{eq_f_tau_c} as the convolution implicitly assumes the \ac{MU} starts at a random position inside the inner circle.}
This relation is evaluated numerically as no closed-form expression exists for the $N$-fold convolution of any of the functions in Eqs. \eqref{eq_f_tau_c}.
To summarize, we evaluate the probability of the \ac{MU} leaving the inner circle for the first time with \eqref{eq_P_tau_j}, where $P_{J\mathrm{out}}(j)$ as given by \eqref{eq_first_time_discrete} and $P(T_\mathrm{out}|j)$ is evaluated numerically by integrating the \ac{PDF} in \Cref{eq_f_tau_c}.

\subsection{Validating the Theoretical Expression for $\mathrm{P}_\mathrm{in}(t)$}
\label{sec_validation}

We validated the theoretical expressions for the $\mathrm{P}_\mathrm{in}(t)$ in~\eqref{eq_Pr_in} through simulations.
We evaluated the probability after \num{1000} realizations accounting for the time instant when the \ac{MU} leaves the inner circle for the first time.
We simulated the three mobility models tracking the position of the \ac{MU} and recorded the time when the \ac{MU} leaves the inner area, giving the results illustrated in \Cref{fig_Pr_in_sim} for the three different models in \Cref{eq_f_tau_j_2,eq_f_tau_j_3,eq_f_tau_j_4}.
The theoretical formulation for $\mathrm{P}_\mathrm{in}$ in \eqref{eq_Pr_in} manifests a good correspondence with the simulation of the three \ac{RWP} mobility models as illustrated in \Cref{fig_Pr_in_sim}.

\subsection{Evaluating the Probability that the {MU} Stays Outside of the Illuminated Area}
\label{sec_Pr_out}

To evaluate the probability that the \ac{MU} stays outside the illuminated area, we follow the same procedure as in \Cref{sec_Pr_in}.
We compute the first-hitting time at which the \ac{MU} reaches the inner circle $P_\mathrm{fh\_in}$ when initially being in the outer area, and evaluate the probability for the \ac{MU} to remain in the outer circle as
\begin{equation}\label{eq_Pr_out}
    \mathrm{P}_\mathrm{out}(t)=1-P_\mathrm{fh\_in}(t).    
\end{equation}
To compute $P_\mathrm{fh\_in}$, we evaluate the survival probability as in \Cref{eq_first_time_discrete}, just replacing $A_\mathrm{in}$ with $(A-A_\mathrm{in})$.
Then, to evaluate $P_\mathrm{fh\_in}(t)$, we use the same formulation for the total probability as in \eqref{eq_P_tau_j}, but replacing $T_\mathrm{out}$ with $T_\mathrm{in}$.
Besides, $P(T_\mathrm{in}|j)$ is evaluated with its corresponding \ac{PDF}, similarly formulated as in \eqref{eq_f_tau_c}, after replacing $r_\mathrm{in}$ with $r_\mathrm{out}$, to evaluate $f_{\tau_j;r_\mathrm{out}}$.
We note that, following these calculations, we implicitly assume the worst-case scenario, thus avoiding the calculation of possible interceptions between the \ac{MU} trajectory and the illuminated area.
\section{{AoI} Framework}
\label{sec_AoI}

In the following, we theoretically formulate the average \ac{PAoI} metric to balance the overhead and data transmission time slots.
We aim to evaluate the update period of the \ac{IRS}, with the value of $T_\mathrm{upd}$ in \Cref{fig_transmission}, and balance the trade-off between overhead and transmission opportunities.
The more frequently we update the \ac{IRS} (smaller $T_\mathrm{upd}$), the higher the chances of error-free transmissions, as the \ac{MU} will be mostly in the illuminated area. 
However, with the increased update frequency, the overhead becomes larger, thereby limiting the availability of time slots to transmit data.
Conversely, reducing the frequency of updates (increased $T_\mathrm{upd}$) allows more time slots for transmission.
Yet, this is at the expense of more errors and less throughput, as the \ac{MU} will increasingly spend more time outside the illuminated area between \ac{IRS} updates.
Such a dynamic behavior evidences a trade-off between overhead and throughput that we aim to balance with the \ac{AoI} framework in this section.

\begin{figure}
\centering
\includegraphics[width=0.8\linewidth]{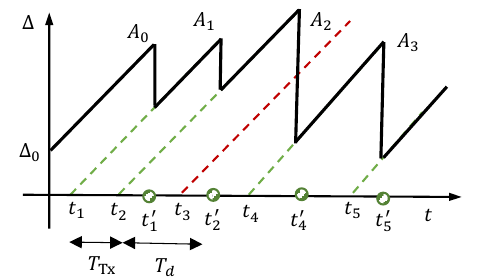}
\caption{Illustration of the \ac{AoI} of packets (dashed lines) and as perceived on a given destination node (bold line).}%
\label{fig_AoI}
\vspace{-.8em}
\end{figure}

The \ac{PAoI} metric results from the sequence of peaks in the \ac{AoI} curve, which is evaluated based on the generation ($t_{i+1}-t_i$) and system ($t_i'-t_i$) times, where $t_i$ denotes the time instant when a packet is generated, and $t'_i$ is the time when it is successfully received; see their representation in \Cref{fig_AoI}.\footnote{This definition follows~\cite{pappas2023age} and \cite[Def. 3.1]{kosta2017age}.}
Following the illustration in this figure, the \ac{AoI} curve is the continuous line and the \ac{PAoI} refers to the sequence of peaks~$A_i$.
The \ac{AoI} curve will linearly increase between consecutive receptions (e.g., between $t_1'$ and $t_2'$) following the increasing age of the most recent received packet (e.g., at $t_1'$).
As a result, the \ac{AoI} is a sawtooth curve with peaks $\{A_0,\, A_1,\dots\}$ representing the highest age at the reception time. 
The average \ac{PAoI}, denoted as~$\Delta^{(p)}$, corresponds to the average of the sequence of peaks and is evaluated as
\begin{equation}\label{eq_def_PAoI}
 \Delta^{(p)}=\mathbb{E}[A_i]=\mathbb{E}[T_g]+\mathbb{E}[T_s]
\end{equation}
as follows from~\cite[Def. 3.3.2]{kosta2017age}, where $T_g=t_{i+1}-t_i$ denotes the generation time, and $T_s=t'_i-t_i$ is the system time.

\subsection{Integrating the System Model within the
{AoI} Concept}

Although \Cref{eq_def_PAoI} is not that explicit \textit{per se}, the height of the peaks in the \ac{AoI} directly reflects the dynamics in the communication link.
We substantiate this statement through the following facts:
%
\begin{itemize}
    \item The transmission time in the downlink ($T_\mathrm{Tx}$ in \eqref{eq_T_Tx}) corresponds to the time interval between the generation of two successive packets in the \ac{AoI} formulation, i.e., the generation time~\mbox{($t_{(i+1)}-t_i$)} is equal to $T_\mathrm{Tx}$; see the illustration in \Cref{fig_AoI}.
    Besides, as~$T_\mathrm{Tx}$ also accounts for the communication overhead and the latency (due to the packet size) in the link, the generation time also conveys the same two components (overhead and latency); see the dependency of~$T_\mathrm{Tx}$ with $T_\mathrm{ovh}$ and $T_\mathrm{data}$ in \eqref{eq_T_Tx}.
    %
    \item The two link conditions, error-free transmission and transmission with errors, are included within the system time $(T_s=t'_i-t_i)$.
    When the link is error-free, the system time is the minimum possible and equal to the transmission delay, e.g., $t_2'-t_2=T_d$; see \Cref{fig_transmission}.
    However, when the link causes errors, the system time increases until the \ac{MU} returns to the area illuminated by the \ac{IRS} or the \ac{IRS} is again reconfigured to illuminate the \ac{MU}.
    In this regard, the system time is a random variable that depends on the size of the illuminated area (defined by the \ac{IRS}), the mobility pattern of the \ac{MU}, and the update period of the \ac{IRS}.
    \item As a result of the above interrelation of the generation and system times with the system model parameters, the impact of the transmission overhead, \ac{IRS} reconfiguration, and mobility is jointly evaluated within the amplitude of the peaks in the \ac{AoI} curve, as directly follows from \Cref{eq_def_PAoI}.
    In this way, the average \ac{PAoI} metric accounts for the overhead and the dynamics of the communication link between interruption of transmissions, errors-free transmission, and transmission with error.
    \item Finally, the average \ac{PAoI} metric is minimized based on the \ac{IRS} update period ($T_\mathrm{upd}$), allowing us to optimally balance the impact of overhead and throughput in the link.
\end{itemize}
In the next section, we elaborate on a closed-form expression for the average \ac{PAoI} metric and illustrate the optimal \ac{IRS} update period.

\subsection{Closed-form expression for the average \ac{PAoI}}

To obtain a closed-form expression for $\Delta^{(p)}$, we consider the temporal evolution of the \ac{AoI} curve in \Cref{fig_AoI}.
The evolution of the \ac{AoI} curve follows the dynamic of the communication system according to the three states illustrated in \Cref{fig_markov_chain}.
The corresponding height of the peaks can be evaluated for these three different cases as follows:
\begin{enumerate}[label=\textbf{{Case~\arabic*}}, labelindent=0pt, wide, labelwidth=!]
\item (\ac{IRS} is reconfigured): This is the initial state, where the \ac{MU} is localized and the \ac{IRS} is reconfigured; see \Cref{fig_transmission}.
In this state, transmissions are interrupted along the time interval $T_\mathrm{conf}=T_\mathrm{loc}+T_\mathrm{IRS}$.
This case is already included within the transmission time interval $T_\mathrm{Tx}$ in \Cref{eq_T_Tx}; see the dependency of the $T_\mathrm{ovh}$ term with ($T_\mathrm{loc}+T_\mathrm{IRS}$) in \Cref{eq_ovh}.
\item (\ac{MU} is inside the illuminated area):
In this case, errors do not occur, and the height of the peaks can readily be evaluated as
\begin{equation}\label{eq_age_in}
A_{i,\mathrm{in}}=T_d+T_\mathrm{Tx},
\end{equation}
along the time interval inside the illuminated area, where \mbox{$T_d=\frac{c}{d}$} is the communication delay between the \ac{MU} and the \ac{AP}, $c$ is the speed of light and $d$ is the sum of the distances between the \ac{MU} and the \ac{IRS} and between the \ac{IRS} and the \ac{AP}.
The relation in \eqref{eq_age_in} computes the time elapsed between two consecutive receptions~\cite{costa2016age};
see for example the reception at $t'_1$ and $t'_2$ in \Cref{fig_AoI}, where \mbox{$T_\mathrm{Tx}=t_{i+1}-t_i$} and $T_d=t'_i-t_i$ are assumed to be deterministic variables.
This event (amplitude of the peaks) will occur with the steady state probability $\pi_j|_{j=2}$ as a result of evaluating \eqref{eq_trans_prob2}.
\item (\ac{MU} is outside the illuminated area): In this case, packets are not decoded and thereby dropped, leading to an increase of the \ac{AoI}.
See the case depicted in \Cref{fig_AoI}, where the packet at $t_3$ is not received, and the age increases till the reception of the next emitted packet at~$t'_4$.
Specifically, the peak \ac{AoI} increases by the amount $T_\mathrm{Tx}$, i.e., as~\mbox{$(T_\mathrm{Tx}+T_\mathrm{data}+T_\mathrm{Tx})$}, as the receiver must wait to create the next packet at $t_4$.
%
When considering an arbitrary time interval for the \ac{MU} to stay outside the illuminated area, denoted as~$T_\mathrm{o}$, the waiting time before the next packet arrives yields~\mbox{$\left\lceil\frac{T_\mathrm{o}}{T_\mathrm{Tx}}\right\rceil\times T_\mathrm{Tx}$}.
In this way, the corresponding peak in the \ac{AoI} curve is readily evaluated as
\begin{equation}\label{eq_age_out}
A_{i,\mathrm{out}}=T_d+T_\mathrm{Tx}+\left\lceil\frac{T_\mathrm{o}}{T_\mathrm{Tx}}\right\rceil T_\mathrm{Tx},  
\end{equation}
occurring with the steady state probability $\pi_j|_{j=3}$ as a result of evaluating \Cref{eq_trans_prob2}. 
This formulation implicitly assumes that errors will happen independently of the time duration outside the illuminated area, i.e., as long as $T_\mathrm{o}>0$.
\end{enumerate}

We evaluate the average \ac{PAoI} based on Cases \num{2} and~\num{3}, as we consider the overhead from the first case already introduced in the formulation for $T_\mathrm{Tx}$ as in \eqref{eq_T_Tx}. 
For this evaluation, we use the fraction of time relative to the time interval for transmissions only, i.e., after the localization and reconfiguration phases along the time interval $(T_\mathrm{upd}-T_\mathrm{loc}-T_\mathrm{IRS})$; see \Cref{fig_transmission}.
Accounting for the inclusion of $T_\mathrm{conf}=T_\mathrm{loc}+T_\mathrm{IRS}$ within the Cases \num{2} and~\num{3}, we proportionally increase both probabilities $\pi_2$ and $\pi_3$ with $(\pi_2\pi_1)$ and $(1-\pi_2)\pi_1$, respectively, as these terms evaluate the fraction of time (given by $\pi_2\pi_1$ in Case 2) of transmissions that proportionally includes $T_\mathrm{conf}$, as provided by $(\pi_2+\pi_2\pi_1)T_\mathrm{upd}$.
These evaluations yield the following probabilities
\begin{align}\label{eq_pi_2_3}
    \pi_2'&=\pi_2+\pi_2\pi_1\\
    \pi_3'&=\pi_3+(1-\pi_2)\pi_1\nonumber,
\end{align}
complying with the relation $\pi'_2+\pi'_3=1$.
Then, using \eqref{eq_age_in} and \eqref{eq_age_out} and evaluating the average time that the \ac{MU} is outside the illuminated area as $T_\mathrm{o}=\pi_3T_\mathrm{upd}$, we can calculate the average \ac{PAoI} as
\begin{align}\label{eq_PAoI}
    \Delta^{(p)}&=A_\mathrm{in}\times\pi'_2+A_{i,\mathrm{out}}\big|_{T_\mathrm{o}=\pi_3T_\mathrm{upd}}\times\pi'_3,\\
    &=(T_d+T_\mathrm{Tx})\times(\pi'_2+\pi'_3)+\left\lceil\frac{\pi_3T_\mathrm{upd}}{T_\mathrm{Tx}}\right\rceil T_\mathrm{Tx}\times\pi'_3, \nonumber\\
    &=(T_d+T_\mathrm{Tx})+\left\lceil\frac{\pi_3T_\mathrm{upd}}{T_\mathrm{Tx}}\right\rceil T_\mathrm{Tx}\times\pi'_3. \nonumber
\end{align}

\begin{figure}
    \centering
    \includegraphics[width=0.8\linewidth]{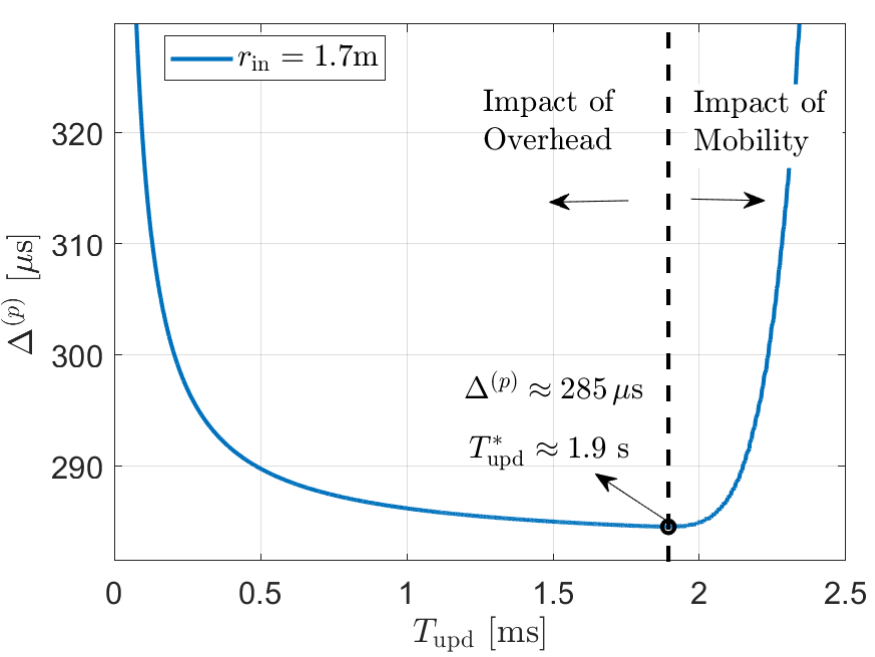}
    \caption{Resulting average \ac{PAoI} assuming the \ac{RWP} mobility model with $v=\SI{1}{\meter\per\second}$ for the \ac{MU}, $r_\mathrm{in}=\SI{1.7}{\meter}$, and outer radius $r_\mathrm{out}=\SI{3}{\meter}$.
    }
    \label{fig_AoI_3m}
    \vspace{-.8em}
\end{figure}

\Cref{eq_PAoI} jointly accounts for the impact of the overhead and communication errors.
The overhead is included in $T_\mathrm{Tx}$ (see \eqref{eq_T_Tx}), while the communication errors are accounted for in the terms $\pi_3$ and $\pi'_3$, as these two probabilities reflect the case when the \ac{MU} is outside the illuminated area.
To illustrate these two mechanisms, we depict the impact of the two mentioned components in \Cref{fig_AoI_3m}.
Reducing $T_\mathrm{upd}$ will increase the \ac{AoI} as the overhead increases.
Conversely, increasing $T_\mathrm{upd}$ will also increase the age of information as the \ac{MU} will spend time outside the illuminated area with increasing probability.
In the next section, we provide more insights into this calculation and its practical use to balance the overhead introduced by the \ac{IRS} in the communication link.

\section{Results: Balancing the impact of the overhead and packet losses for a WiFi scenario}
\label{sec_results}

This section illustrates the application of the above theoretical formulation and studies the optimal reconfiguration periodicity for the \ac{IRS}.
The optimal balance is attainable by minimizing the expression in \Cref{eq_PAoI}; however, its evaluation is not straightforward as we need first to evaluate its dependency on the probabilities $\pi_2',\, \pi_3'$ and $\pi_3$ within the application scenario.
We also need to evaluate the dependency of the average \ac{PAoI} metric with $T_\mathrm{upd},\,T_\mathrm{Tx}$ and $T_d$ according to the communication protocol and the communication distances.
For this evaluation, we choose a functional scenario where we implement a communication link over WiFi in the mmWave band.\footnote{Other particular communication scenarios (as in \ac{5G}) are also applicable as long as they follow a packet-format scheme as illustrated in \Cref{fig_transmission}.}
We illustrate the average \ac{PAoI} metric for a communication system compliant with the IEEE 802.11ad standard and provide realistic performance values.

%

\subsection{Scenario Setup}
\label{sec_scenario}
We follow the IEEE 802.11ad standard as it operates in the mmWave band \cite{nitsche2014ieee} through the schematic in \Cref{fig_location}.
The communication link consists of a single \ac{AP}-\ac{MU} link through the \ac{IRS}-assisted channel; where we assume \ac{NLOS} conditions between the \ac{AP} and the \ac{MU}.
In this communication link, we locate the \ac{AP} at $\SI{2.5}{\meter}$ height from the floor, and \ac{MU}'s terminal is located at $\SI{1.5}{\meter}$ height.
The communication between the \ac{AP} and the \ac{MU} occurs through the \ac{IRS} located in the ceiling at a height of $\SI{3}{\meter}$, where the \ac{MU} is within the near field region of the \ac{IRS}.

The \ac{AP} generates a random data sequence and creates an IEEE 802.11ad packet with the preamble and data field length, as listed in \Cref{tab_parameters}.\footnote{For details on the IEEE 802.11ad packet structure, we refer to the \ac{DMG} PHY implementation, as given within the standard in \cite[Sec. 20]{ieee80211-2020}. 
A summary of these parameters is also listed in~\cite{matlabWLANPPDU}.}
We perform transmissions with the \ac{MCS} \num{12.6}, which corresponds to a single-carrier 64-\ac{QAM}, encoded with a \ac{LDPC} code; see \cite[Table 20.19 and Sec. 20.6.3.1.4]{ieee80211-2020}.\footnote{$\mathrm{MCS}=12.6$ enables high throughput transmissions (\SI{8}{\giga\bit\per\second}) and large data packets ($\approx \SI{262}{\kilo\byte}$ per frame) supporting applications like 3D high-definition video streams.
This setting affords transmissions in long-lasting channel coherence times; which is the case in our scenario where the \ac{MU} speed is not larger than $\SI{1}{\meter\per\second}$.}

\begin{figure}
 \centering
 \includegraphics[width=0.8\linewidth]{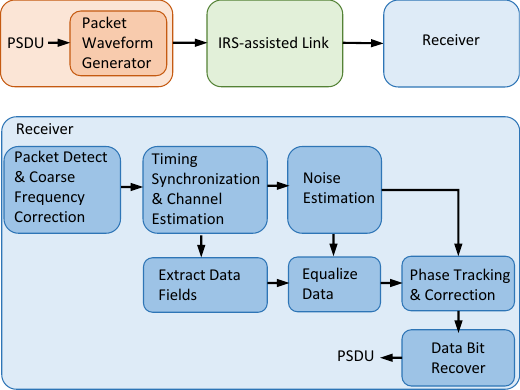}
 \caption{Block diagram for the implemented transmission-reception scheme according to the 802.11ad.}
 \label{fig_Tx_Rx_scheme}
\vspace{-0.8em}
\end{figure}

As illustrated in \Cref{fig_location}, the signal travels from the \ac{AP} to the \ac{IRS}, and from there, it is reflected to the ground plane where the \ac{MU} is located.
We abstract the communication link with the \ac{SNR} parameters as evaluated in \Cref{eq_SNR}, where $\lambda=\frac{c}{f_c}\approx5\si{\milli\meter}$ corresponds to the center frequency of transmissions $f_c=\SI{60.48}{\giga\hertz}$.
This transmission frequency is the selected carrier frequency for channel two within the 802.11ad standard and is available in all regions.
The receiver decodes the information bits in the data field after various signal processing blocks accounting for packet detection, time and frequency synchronization, and channel and noise estimation.
See details in the block diagram representation of the link and the receiver in \Cref{fig_Tx_Rx_scheme}.
The transmission and noise power levels at the receiver are selected from the ETSI report in \cite{etsi_ris003-v111} and listed in \Cref{tab_parameters}.


\begin{table}
  \centering
  \caption{Simulation parameters.}
        \begin{tabular}{rlll}
    \toprule
          & Variable & Description & Value \\
    \midrule
    \multicolumn{1}{c}{\multirow{6}[2]{*}{\begin{sideways}Frame parameters\end{sideways}}} &       & \multicolumn{1}{p{11em}}{Preamble field length} & \num{4352} samples \\
          &       & \multicolumn{1}{p{11em}}{Maximum data field length} & \num{456768} samples\\
          &       &       & \multicolumn{1}{p{4.045em}}{$\approx~\SI{262}{\kilo\byte}$} \\
          &       & \multicolumn{1}{p{11em}}{Training field length} & \num{3712} samples \\
          &       & Idle time & $\SI{20}{\micro\second}$ \\
          & \acs{MCS} & \multicolumn{1}{p{11em}}{Modulation and coding scheme} & \num{12.6} \\
    \midrule
          & $N$   & \multicolumn{1}{p{11em}}{Total of IRS reflecting elements} & $160 \times 160$ \\
    \midrule
    \multicolumn{1}{c}{\multirow{11}[2]{*}{\begin{sideways}\acs{RF} parameters\end{sideways}}} & $f_s$ & {Sampling rate} & $\SI{1.76}{\giga\hertz}$ \\
          & $\mathrm{BW}$ & {Receiver bandwidth} & $\SI{2640}{\mega\hertz}$ \\
          & $f_c$ & \multicolumn{1}{p{11em}}{Carrier frequency, corresponding to Channel 2} & $\SI{60.48}{\giga\hertz}$ \\
          & $P_\mathrm{Tx}$ & \multicolumn{1}{p{11em}}{Transmitter peak power} & $\SI{30}{\decibelm}$ \\
          &       & {Number of \ac{AP} antennas} & \num{8} \\
          &       & \multicolumn{1}{p{11em}}{Number of \ac{AP} beams} & \num{4} \\
          &       & Gain per AP beam & $\SI{9}{\decibeli}$ \\
          & $\mathrm{NF}$ & Noise figure & $\SI{7}{\decibel}$ \\
          & $\mathrm{N}_0$ & {Thermal noise density} & $\SI{-174}{\decibelm\per\hertz}$ \\
          & $\mathrm{N}_\mathrm{int}$ & \multicolumn{1}{p{11em}}{Receiver interference density} & $\SI{-165.7}{\decibelm\per\hertz}$ \\
          & $P_\mathrm{N}$ & Noise floor & $\approx \SI{-70}{\decibelm}$ \\
    \midrule
          & $\mathbf{p}_\mathrm{AP}$ & \ac{AP}'s location & $[2,\, 0,\,2.5]\, [\si{\meter}]$ \\
          & $\mathbf{p}_\mathrm{IRS}$ & \ac{IRS}'s location & $[2,\, 3,\,3]\, [\si{\meter}]$ \\
          & $\mathbf{p}_\mathrm{MU}$ & \ac{MU}'s initial location & $[2,\, 3,\,1.5]\, [\si{\meter}]$ \\
          & $v$   & \ac{MU}'s speed & $\SI{1}{\meter\per\second}$ \\
    \bottomrule
    \end{tabular}%
    \begin{tablenotes}
     \item The parameters for the length of the preamble, data field, and training subfield follow the IEEE 802.11ad PHY frame parameters and are quantified within the Matlab code in \cite{802_11_IRS}.
     The preamble field encompasses the short training, channel estimation; and header fields.
     The training subfields encompass the automatic gain control and beamforming training fields; see \cite{matlabWLANPPDU}.
     \item We follow the ETSI report in \cite[Table 5, entries (13) to (17) pp. 51]{etsi_ris003-v111} to define the noise-related values.
     We evaluate the noise floor as \mbox{$P_\mathrm{N}=10\log(10^{\frac{\mathrm{NF}+N_o}{10}}+10^{\frac{N_\mathrm{int}}{10}})+10\log(\mathrm{BW})$}.
     Regarding the \ac{IRS}, we choose $160 \times 160$ elements, equivalently to a size of \mbox{$160\times \frac{\lambda}{2}=160\frac{c}{2f_c}\approx\SI{40}{\centi\meter}$}, where $c$ is the speed of light.
     We choose this value as we observe a sufficiently high \ac{SNR} in the center; see \Cref{fig_IRS_illu}.
     An example of fabricated solutions is $160\times 160$ in the $\SI{60}{\giga\hertz}$ band~\cite{kamoda2011ghz}. 
    \end{tablenotes}
  \label{tab_parameters}%
  \vspace{-.8em}
\end{table}%

%

\begin{figure}
 \centering
 \includegraphics[width=0.8\linewidth]{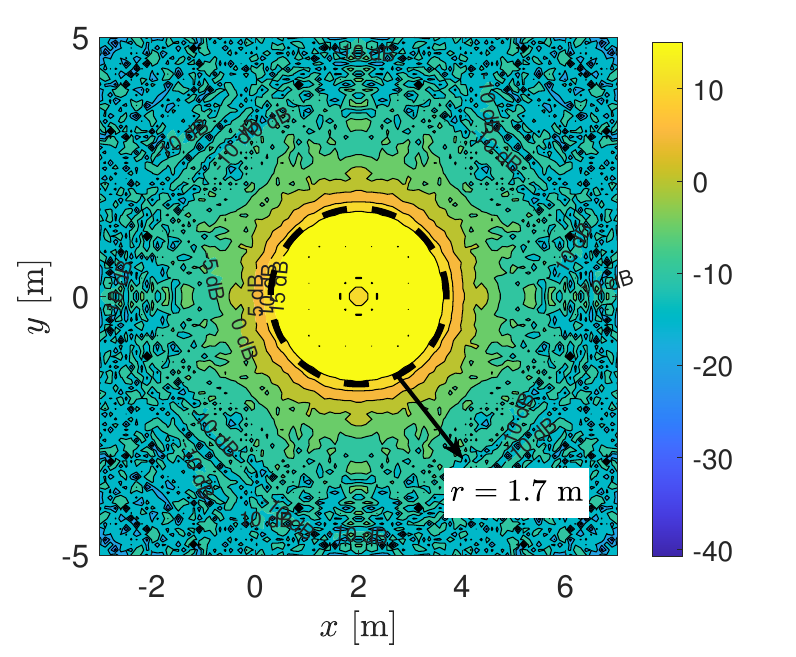}
 \caption{Heat map of the \ac{SNR} values for the \ac{IRS}'s illuminated area and the parameters given in \cref{tab_parameters}, following the framework presented in \cite{alexandropoulos2022nearfield}.}
 \label{fig_IRS_illu}
\vspace{-.8em}
\end{figure}

We configure the \ac{IRS} of $160 \times 160$ elements as indicated in \Cref{sec_IRS}.
We selected the \ac{IRS} dimension as follows from the fabricated module in \cite{kamoda2011ghz}.
To ensure seamless communication for the \ac{MU}, we configure the \ac{IRS} to render an \ac{SNR} larger than $\SI{30}{\decibel}$ within the illuminated area.
There, we observe zero \ac{BER} for encoded transmissions.
Following this setup, the \ac{IRS} projects a nearly circular \ac{SNR} pattern on the ground since it is installed in the ceiling; see \Cref{fig_IRS_illu}.
The results presented in \cref{fig_IRS_illu} assure the largest illuminated area for the corresponding \ac{SNR} value larger than $\SI{30}{\decibel}$ with a wide-beam parameter of \mbox{$\Delta x=\Delta y=1.7$ $\si{\meter}$}.
As depicted in \cref{fig_IRS_illu}, outside the illuminated area, the \ac{MU} experiences lower \ac{SNR} values, as the beam from the \ac{AP} is steered by the \ac{IRS} mainly to inside the area.

\subsection{Evaluating the Transmission Frame Parameters}

We evaluate the variables $T_\mathrm{IRS}$, $T_\mathrm{loc}$, $T_\mathrm{p\_ovh}$, and $T_\mathrm{data}$ using the parameters listed in \Cref{tab_parameters} as follows:

\noindent
\textbf{\ac{IRS} reconfiguration} ($T_\mathrm{IRS}$): We assume the \ac{IRS} is equipped with a WiFi interface, and the \ac{AP} sends the pre-calculated \ac{IRS} coefficients, as introduced in \Cref{sec_scenario}, through this interface.
Although the coefficients can be calculated at the \ac{IRS}, we select to be evaluated at the \ac{AP} in favor of reduced computational complexity with the \ac{IRS} equipment.
In this setup, we also assume that the reconfiguration time equals the transmission time of a WiFi packet, where the data block within the frame is of the minimum length needed to send the coefficients from the \ac{AP} to the \ac{IRS}.
Assuming a double point precision of~\num{32} bits per coefficient, and that the \ac{IRS} comprises $160\times 160$ coefficients (see \Cref{tab_parameters}), the amount of bytes needed is $\mathrm{Data}_\mathrm{IRS}=\left\lceil\frac{32\times 160^2}{8}\right\rceil=\SI{102.4}{\kilo\byte}$ yielding
    \begin{align}\label{eq_T_IRS}
        T_\mathrm{IRS}&=T_\mathrm{PPDU-IRS},\\
        =&\frac{1}{f_s}(\text{Preamble field length + Data}_\text{IRS} \nonumber\\
        &\text{ + Training field length}),\nonumber\\
        &\approx \SI{105.4}{\micro\second},\nonumber
    \end{align}
    %
This calculation follows the \ac{PPDU} structure, according to the IEEE 802.11ad standard (see details in \cite{matlabWLANPPDU}).
The preamble, training lengths, and sampling frequency $f_s$ are specified in \Cref{tab_parameters} while $\text{Data}_\text{IRS}$ is evaluated as already indicated above.\footnote{The corresponding field length is evaluated through the Matlab code accessible in \url{https://github.com/tkn-tub/IRS_Rejuvenation}}    
    
\noindent
\textbf{Localization time} ($T_\mathrm{loc}$):
    We follow the solution reported in \cite{zhang2023multiperson} for localization, where an \ac{IRS} is reconfigured to scan the \ac{MU} plane and look for its reflected signals in a WiFi scenario.\footnote{During the localization phase, the \ac{IRS} is configured differently than for data transmission as it renders a narrow beam instead of a wide beam (as explained in \Cref{sec_IRS}).
    We omit details on the \ac{IRS} reconfiguration for localization as this step is abstracted with the localization time variable; still, information on this method is provided in \cite[Algorithm 1]{zhang2023multiperson}.}    
    The method performs once a coarse-grained measurement to obtain the initial position of the \ac{MU} and continues tracking the user with fine-grained measurements.
    The fine-grained measurement is the one we account for with the localization time, as the coarse-grained measurement is performed only once, yielding negligible associated overhead over time.
    For the fine-grained measurement, a \mbox{$1\times 1$ \si{\meter^2}} area around the \ac{MU} location is divided into $11\times 11$ blocks; see \cite[Sec. VI A]{zhang2023multiperson}.
    The \ac{IRS} is reconfigured once per block by the \ac{AP}, and then measurements are performed with the \ac{AP} sending two data packets per block via the \ac{IRS}.
    Adopting this approach and setup, the localization time yields
    %
    \begin{align}\label{eq_T_loc}
        T_\mathrm{loc}=&(2\text{ transmissions})\times (11^2 \text{ blocks})\times T_\mathrm{PPDU-loc}+\\
        &+(11^2 \text{ blocks})\times T_\mathrm{PPDU-IRS} \nonumber\\
        \approx& \SI{15}{\milli\second}, \nonumber
    \end{align}
where the first term accounts for the transmission of two packets per block, and the second term for the \ac{IRS} reconfiguration per block.
The parameter $T_\mathrm{PPDU-IRS}$ is calculated similarly as in \eqref{eq_T_IRS} but with $\text{Data}_\text{loc}=1$, which refers to the transmission of a single byte within the data block in the frame.\footnote{During localization, we assume a single transmitted byte per packet to afford the lowest overhead possible.}
We remark that \Cref{eq_T_loc} only reflects time slots for packet transmissions, as we assume that the computation time required to run the localization algorithm is negligible when compared to the transmission time.
The implicit assumption is that the computational capabilities at the \ac{AP} are sufficiently high.



\noindent
\textbf{Preamble and training time} ($T_\mathrm{p\_ovh}$): The preamble field of the received packet is used to compensate for the channel effects and equalize data (frequency correction, noise estimation, synchronization), also for beamforming (training field), and further parameter specification like the \ac{MCS} within the header block.
    We calculate the overhead within the packet as
    \begin{align}
        T_\mathrm{p\_ovh}&=\frac{1}{f_s}(\text{Preamble field length + Training field length}),\\
        &\approx \SI{5.3}{\micro\second}\nonumber.
    \end{align}
    
\noindent
\textbf{Data time interval} ($T_\mathrm{data}$): We compute this time interval with the total of samples for the data sequence as
    \begin{align}
        T_\mathrm{data}&=\text{Data Length}/ f_s\\
        &\approx \SI{260}{\micro\second},\nonumber
    \end{align}
    where we assume the largest packet possible (corresponding to around $\SI{262}{\kilo\byte}$) within this standard, and in favor of the least packet overhead.


%

\subsection{Evaluating the Average PAoI}

\begin{figure}
    \centering
    \includegraphics[width=0.8\linewidth]{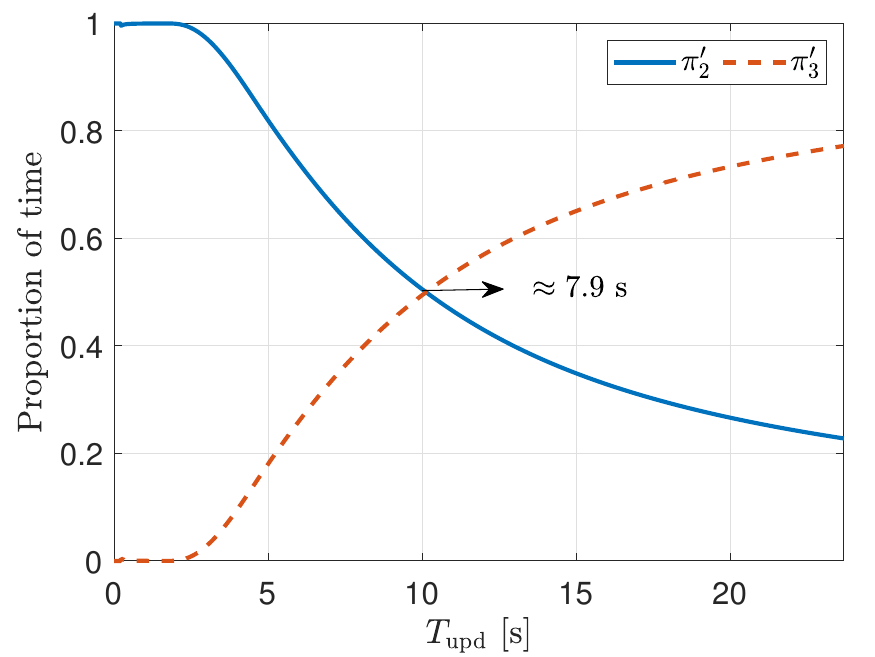}
    \caption{Fraction of time in the inner and outer areas when $r_\mathrm{in}=\SI{1.7}{\meter}$, $r_\mathrm{out}=\SI{3}{\meter}$, $v=\SI{1}{\meter\per\second}$.}
    \label{fig_pi_2_3_v2}
    \vspace{-.8em}
\end{figure}

\begin{figure}
 \centering
 \includegraphics[width=0.8\linewidth]{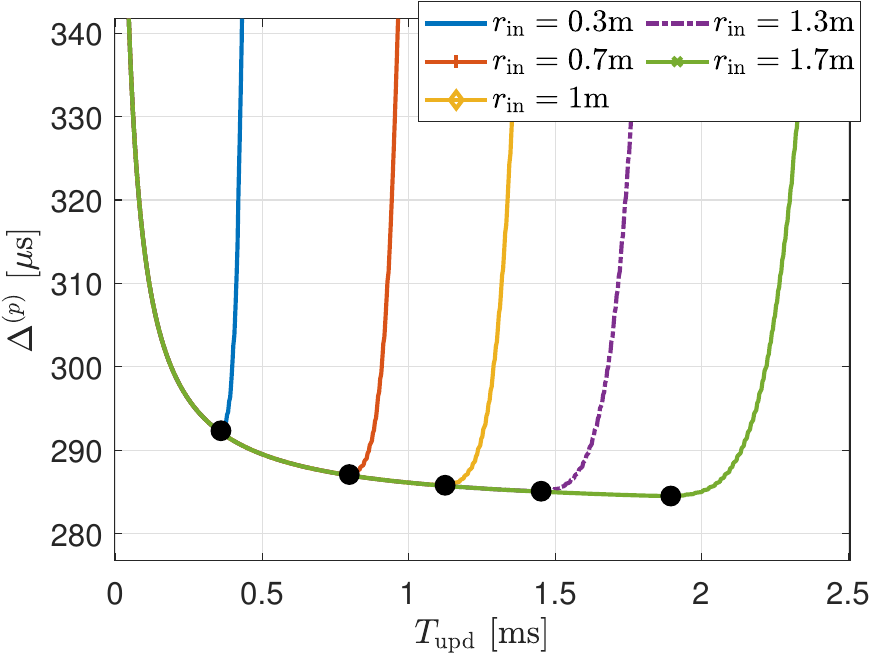}
 \caption{Average \ac{PAoI} with $T_\mathrm{upd}$ and various radius for the illuminated area.}
 \label{fig_PAoI_results}
\vspace{-.8em}
\end{figure}

Using the IEEE 80211.ad transmission-reception scheme, we balance the impact of overhead (due to \ac{MU} localization and \ac{IRS} reconfiguration) and packet losses (due to \ac{MU} mobility) through the following steps
\begin{enumerate}[label=\textbf{\textbf{Step~\arabic*.}}, labelindent=0pt, wide, labelwidth=!]
    \item \textbf{(Evaluate the radius for the illuminated area):} We compute this radius to observe a zero \ac{BER} as the criteria.
    \footnote{We choose this criteria as our formulation for the average \ac{PAoI} in \eqref{eq_PAoI} implicitly assumes free-of-errors transmissions in the illuminated area.
    To consider cases allowing higher \ac{BER}, a new formulation is needed to evaluate the \ac{PAoI} to consider communication errors.}
    The \ac{BER} is evaluated using Monte Carlo simulations of the transmission-reception scheme depicted in \Cref{fig_Tx_Rx_scheme}.
    We performed a total of \num{50e4} packet emissions per radius value, which correspond to more than $\SI{100}{\mega\bit}$ emissions.
    \Cref{fig_PER_ad} illustrates the evaluation of the perceived \ac{SNR} (mean and variance) and \ac{BER} at the receiver side as a function of the radius.\footnote{We remark that the \ac{MU} will be located in the \ac{IRS}'s near field and the \ac{SNR} calculations follow the near field evaluation in \cite{alexandropoulos2022nearfield}.}
    Based on the results in \Cref{fig_PER_ad}, we select $r_\mathrm{in}=\SI{1.7}{\meter}$ as the maximum radius to observe $\mathrm{BER}=0$ as per simulation.\footnote{The evaluation of the \ac{BER} in \Cref{fig_PER_ad} accounts for the selected high-throughput \ac{MCS} parameter, i.e., \num{12.6} as in \Cref{tab_parameters}.
    A less order for this parameters yields a larger radius to observe a zero \ac{BER}.}

\begin{figure}
    \centering
    {\includegraphics[width=0.8\linewidth]{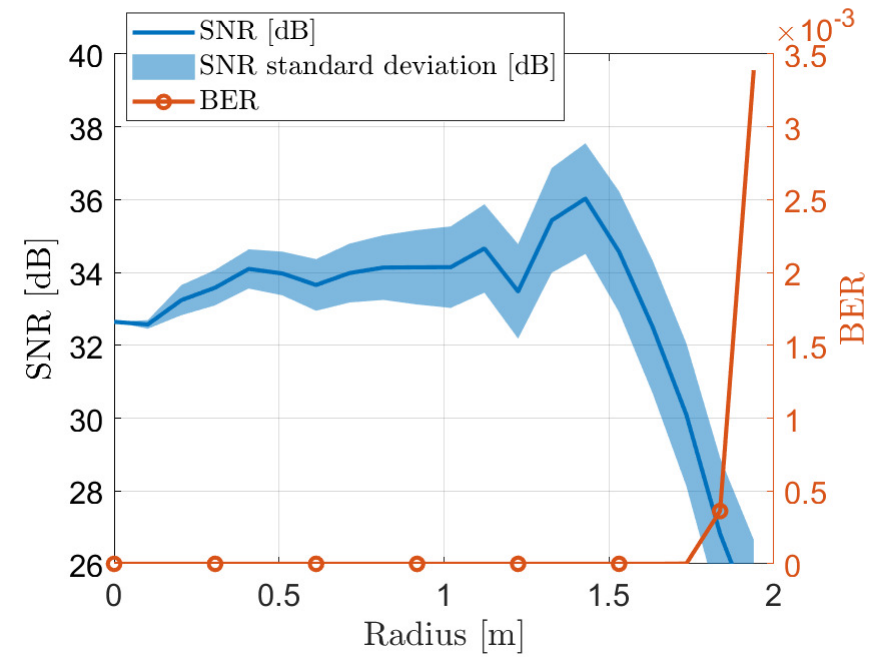}\label{fig_fading_vs_time}}
    \caption{Average \ac{SNR} and \ac{BER} as evaluated along the perimeter of the circle with radius given in the horizontal axis of this plot.}
    \label{fig_PER_ad}
    \vspace{-.8em}
\end{figure}

    \item \textbf{(Evaluate the fraction of time the \ac{MU} is inside and outside the illuminated area):} For a given $T_\mathrm{upd}$, we evaluate the fraction of time the \ac{MU} is inside ($\pi'_2$) and outside~($\pi'_3$) the illuminated area using the formulation in \eqref{eq_pi_2_3}.
    We remark that these fraction of times already include the overhead introduced by the \ac{MU} localization and the \ac{IRS} reconfiguration phase. 
    We show the results of this evaluation in \Cref{fig_pi_2_3_v2}.
    As can be seen, if the \ac{IRS} is updated more frequently than every~$\approx\SI{4.6}{\second}$, the \ac{MU} spends more time in the illuminated area than outside it, and the average \ac{PAoI} will be mostly evaluated as in \eqref{eq_age_in}.

    At this point, we highlight that these two steps capture the influence of the \ac{IRS} design on the communication link.
    For instance, a larger \ac{IRS} effectively expands the illuminated area, thereby shifting to the right the intersection of the curves for \(\pi_2'\) and \(\pi_3'\) in \Cref{fig_pi_2_3_v2}.
    As a result, the overall system performance enhances due to the need to update the \ac{IRS} less frequently.

    %
    \item \textbf{(Evaluate the average \ac{PAoI} and select its minimum):} We evaluate the average \ac{PAoI} with \eqref{eq_PAoI} once the radius of the illuminated area is selected ($r_\mathrm{in}=\SI{1.7}{\meter}$) and the fraction of time has been calculated, see \Cref{fig_pi_2_3_v2}.
    As a result, \Cref{fig_PAoI_results} depicts the average \ac{PAoI} (for various radii) and the optimal update period $T_\mathrm{upd}$.\footnote{For the purpose of illustration only, we evaluate the minimum of the different curves in \Cref{fig_PAoI_results} by selecting the smallest sample in the time sampled range.
    We obtain the samples by evaluating \eqref{eq_PAoI} in the time range \SIrange{0.5e-4}{2.5}{\second} with sampling interval $\SI{50}{\micro\second}$.
   This is an approximate value, and a more accurate solution can be evaluated by derivative-free (e.g., bisection) or gradient-based (e.g., Newton's) methods.}
    With the plot in \Cref{fig_PAoI_results} the average \ac{PAoI} ranges in the hundreds of microseconds, and the optimal update period results in the order of seconds.
    
    Contextualizing these numbers, if we select the optimal update period in \Cref{fig_PAoI_results} when $r_\mathrm{in}=1.7\si{\meter}$, i.e., every~\mbox{$T_\mathrm{upd}=\SI{1.9}{\second}$} (see \Cref{fig_AoI_3m} for more details), the data transmissions between the \ac{AP} and the \ac{MU} are \num{6275} packets in a run before reconfiguring the \ac{IRS} again.\footnote{We do this calculation with the ratio $T_\mathrm{upd}/(T_\mathrm{ppdu}+\text{Idle Time})$, where the Idle time is $\SI{20}{\micro\second}$; see \Cref{tab_parameters}.}
    Reducing $T_\mathrm{upd}$ generates a larger overhead, and increasing it yields a larger \ac{BER}.
    With this update time, the resulting overhead is around $\frac{T_\mathrm{ovh}}{T_\mathrm{upd}}\approx\SI{9.6}{\percent}$ of the update period, after using \Cref{eq_ovh} to evaluate $T_\mathrm{ovh}$.
    This relatively small overhead is the minimum achievable one to balance the impact of the \ac{MU} mobility.
     
\end{enumerate}


Besides, visualizing the behavior for the average \ac{PAoI} for smaller radii of the illuminated area provides some compelling insights. 
As depicted in \Cref{fig_PAoI_results}, the optimal selection for $T_\mathrm{upd}$ shifts to the left with the decreasing radius as expected; however, the minimum average \ac{PAoI} increases but not significantly.
For reduced radii, the selection of the optimal $T_\mathrm{upd}$ becomes crucial, as the average \ac{PAoI} increases dramatically for small deviations out of the optimal point; see, for instance, the case $r_\mathrm{in}=\SI{30}{\centi\meter}$ in \Cref{fig_PAoI_results}.\footnote{The case when $r_\mathrm{in}=\SI{30}{\centi\meter}$ becomes relevant for secure wireless communications (see \cite{yu2020robust}) where smaller coverage areas are preferred and the selection of the \ac{IRS} configuration time becomes more relevant.}




%

\subsection{Extending the Evaluation of the Average PAoI}

\begin{figure}
 \centering
 \includegraphics[width=0.8\linewidth]{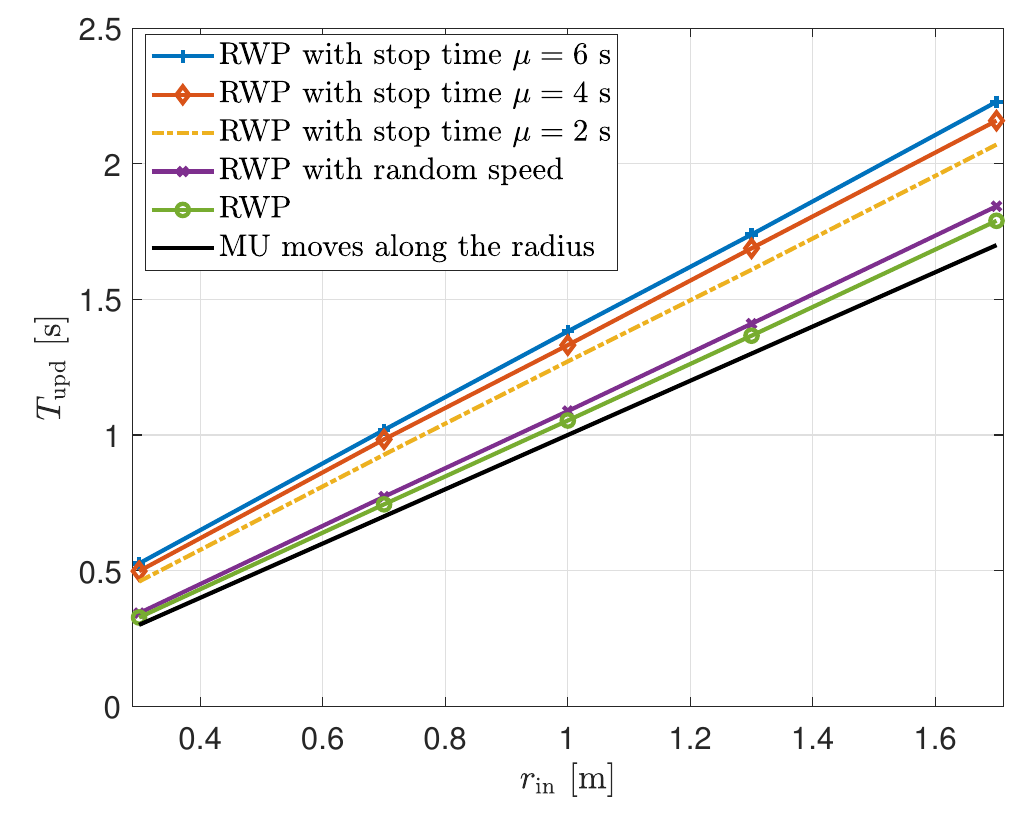}
 \caption{Optimum value of $T_\mathrm{upd}$ versus the inner circle radius for various mobility models.}
 \label{fig_T_upd_vs_r}
\vspace{-.8em}
\end{figure}

Following the above steps, we can evaluate the optimal update periodicity for different mobility model configurations.
\Cref{fig_T_upd_vs_r} comparatively depicts the $T_\mathrm{upd}$ versus the radius of the illuminated area and includes other mobility models such as \ac{RWP} with stop time and with random speed.
For comparisons, we also plot the case when the user moves straight from the center to the perimeter of the illuminated area along the radius.
This represents the worst-case trajectory as the \ac{MU} leaves the illuminated area in the least time.
As for the stop time, we implemented a random variable following an exponential distribution for various average times; see \cite[Section 3.6]{bettstetter2004stochastic}.
As for the random speed, we consider the \ac{MU} changes the speed per waypoint randomly in the range $\SIrange{0.1}{1.5}{\meter\per\second}$ and follows a uniform distribution; see \cite[Section 3.5.1]{bettstetter2004stochastic}.
As \Cref{fig_T_upd_vs_r} illustrates, the relation between the update period and the radius of the illuminated area is almost linear.
In the \ac{RWP} model, $T_\mathrm{upd}$ is slightly larger than the time interval the \ac{MU} takes to travel from the center to the illuminated area's border, i.e., $T_\mathrm{upd}\approx \frac{r_\mathrm{in}}{v}$.
However, this is not the case for \ac{RWP} with stop time where $T_\mathrm{upd}$ results larger (in this case in the range $\SIrange{0.25}{1}{\second}$) than the traveling time from the center to the perimeter of the illuminated area.

\section{Conclusions}
\label{sec_Conclusion}
We presented a methodology to evaluate the optimal update period for the \ac{IRS} elements in mobile scenarios using an average \ac{PAoI} framework.
This methodology allowed us to develop a reconfiguration design with low overhead.
This solution plays a significant role in reducing the communication overhead while preserving the \ac{QoS} of the link.
Given the geometry and the size of the area illuminated by the \ac{IRS}, where the \ac{QoS} criteria are met, our approach evaluates the proportion of time the \ac{MU} will be in the illuminated area to determine later the optimal \ac{IRS} update period.
This procedure minimizes the average \ac{PAoI} and helps jointly balance the impact of overhead and \ac{MU} mobility.
We illustrated the applicability of these calculations considering the practical case of a WiFi link in the mmWave band.
We find that the update period is in the $\si{\milli\second}$ range and is almost linearly increasing with the radius of the \ac{IRS} illumination area.
The resulting minimum information freshness is in the order of the hundreds of $\si{\micro\second}$. 

This paper unlocks several new potential research directions. 
One direction is to extend the \ac{AoI} framework to a multi-link scenario, that is, with multiple \acp{MU} and \acp{IRS}.
In the case of multiple \acp{MU}, the \ac{IRS} need to be split among the \acp{MU}, impacting the achieved \ac{SNR} and the size of the illuminated area.
Another interesting research direction is the optimization of other system parameters, such as the number of \ac{IRS} elements and the transmission policies.
For instance, finding the optimal transmission policy for the \ac{MU} is particularly relevant to avoid transmission while the \ac{MU} is outside the illuminated area, allowing it to extend its battery lifetime.
To this end, a \ac{MDP} can be formulated to determine the optimal transmission policy that jointly maximizes the battery lifetime while minimizing the average \ac{PAoI}.
Moreover, \ac{IRS} splitting could be considered for multiple \ac{MU} scenarios.
This would require finding the optimal policy for jointly updating multiple segments of the same \ac{IRS}.
Another relevant extension is the consideration of more complex channel models for evaluating the average \ac{PAoI} metric.
For instance, including multipath models or small-scale fading will impact the perceived \ac{SNR}, thereby impacting the resulting average \ac{PAoI} metric.
Furthermore, an experimental evaluation of the considered system would allow the validation of the proposed theoretical framework.
Measurement campaigns can be conducted to evaluate the main metrics and correlate their impact on information freshness.

%
\section{Acknowledgments}

We want to thank Dr.\ Anatolij Zubow from TU Berlin for the valuable comments on this manuscript.
This work was supported in part by the Federal Ministry of Education and Research (BMBF, Germany) within the 6G Research and Innovation Cluster 6G-RIC under Grant 16KISK020K.
Jamali’s work was supported in part by the Deutsche Forschungsgemeinschaft (DFG, German Research Foundation) within the Collaborative Research Center MAKI (SFB 1053, Project-ID 210487104) and in part by the LOEWE initiative (Hesse, Germany) within the emergenCITY center [LOEWE/1/12/519/03/05.001(0016)/72].
%

\printbibliography

\end{document}